\documentclass[fleqn,usenatbib]{mnras}

\usepackage{newtxtext,newtxmath}

\usepackage[T1]{fontenc}

\DeclareRobustCommand{\VAN}[3]{#2}
\let\VANthebibliography\thebibliography
\def\thebibliography{\DeclareRobustCommand{\VAN}[3]{##3}\VANthebibliography}

\usepackage{graphicx}	
\usepackage{amsmath}	

\usepackage{lipsum}
\usepackage[english]{babel}
\usepackage{color}
\usepackage[dvipsnames]{xcolor}

\usepackage{multirow}
\usepackage{cleveref}
\usepackage{romannum}
\crefformat{section}{\S#2#1#3} 
\crefformat{subsection}{\S#2#1#3}
\crefformat{subsubsection}{\S#2#1#3}
\usepackage{comment}

\title[The DBL Survey II]{The DBL Survey II: towards a mass-period distribution of double white dwarf binaries}

\author[J.\ Munday et al.]{James Munday,$^{1}$\thanks{Email: james.munday98@gmail.com}
Ingrid Pelisoli,$^{1}$
Pier-Emmanuel Tremblay,$^{1}$
David Jones,$^{2,3}$
Gijs Nelemans,$^{4,5,6}$
\newauthor
Mukremin Kilic,$^{7}$
Tim Cunningham,$^{8\thanks{NASA Hubble Fellow}}$
Silvia Toonen,$^{9}$
Alejandro Santos-Garc\'ia,$^{10}$
Harry Dawson,$^{11}$
\newauthor
Viktoria Pinter,$^{12,13}$
Benjamin Godson$,^{1}$
Llanos Martinez,$^{14}$
Jaya Chand,$^{15}$
Ross Dobson,$^{16}$
Kiran Jhass,$^{17}$
\newauthor
Shravya Shenoy$^{18}$\\
$^{1}$Department of Physics, Gibbet Hill Road, University of Warwick, Coventry CV4 7AL, United Kingdom\\
$^{2}$ Instituto de Astrof\'isica de Canarias, E-38205 La Laguna, Tenerife, Spain\\
$^{3}$ Departamento de Astrof\'isica, Universidad de La Laguna, E-38206, Tenerife, Spain\\
$^{4}$ Department of Astrophysics/IMAPP, Radboud University, P.O. Box 9010, 6500 GL Nijmegen, The Netherlands\\
$^{5}$ Institute for Astronomy, KU Leuven, Celestijnenlaan 200D, 3001 Leuven, Belgium\\
$^{6}$ SRON, Netherlands Institute for Space Research, Niels Bohrweg 4, 2333 CA Leiden, The Netherlands\\
$^{7}$ Homer L. Dodge Department of Physics and Astronomy, University of Oklahoma, 440 W. Brooks St., Norman, OK, 73019 USA\\
$^{8}$Center for Astrophysics, Harvard \& Smithsonian, 60 Garden Street, Cambridge, MA 02138, USA\\
$^{9}$ Anton Pannekoek Institute for Astronomy, University of Amsterdam, 1090 GE Amsterdam, The Netherlands\\
$^{10}$ Departament de F\'isica, Universitat Polit\`ecnica de Catalunya, c/ Esteve Terrades 5, 08860 Castelldefels, Barcelona, Spain\\
$^{11}$ Institute for Physics and Astronomy, University of Potsdam, Karl-Liebknecht-Str. 24/25, 14476 Potsdam, Germany\\
$^{12}$ CTAO, Northen Station,  Rambla Jos\'{e} Ana Fern\'{a}ndez P\'{e}rez 7, ES-38711 Bre\~{n}a Baja, La Palma, Spain\\
$^{13}$ Centro Astron\'omico Hispano en Andaluc\'ia, Observatorio de Calar Alto, Sierra de los Filabres, 04550 G\'ergal, Spain  \\
$^{14}$ Isaac Newton Group of Telescopes, Apartado de Correos 368, E-38700 Santa Cruz de La Palma, Spain\\
$^{15}$ Centre for Advanced Instrumentation, Department of Physics, Durham University, South Road, Durham, DH1 3LE\\
$^{16}$ Mullard Space Science Laboratory, University College London, Holmbury St Mary, Dorking, Surrey RH5 6NT, UK\\
$^{17}$ Department of Physics and Astronomy, University of Sheffield, Sheffield, S3 7RH, UK\\
$^{18}$ Centre for Astrophysics Research, Department of Physics, Astronomy and Mathematics University of Hertfordshire Hatfield, Hertfordshire, AL10 9AB, UK\\}

\date{Accepted 2025 July 18. Received 2025 July 17; in original form 2025 June 1}

\pubyear{2025}

\begin{document}
\label{firstpage}
\pagerange{\pageref{firstpage}--\pageref{lastpage}}
\maketitle

\begin{abstract}
Double white dwarf binaries are an important remnant of binary evolution as they are possible type Ia supernova progenitors and strong sources of gravitational waves in the low-frequency regime. The double-lined double white dwarf (DBL) survey searches for compact double white dwarfs where both stars are spectrally disentangleable. Candidates are identified by being overluminous compared to the cooling sequence of a typical mass, single white dwarf. In this second DBL survey instalment, we present full orbital solutions of 15 double white dwarf binaries from our ongoing campaign to accurately measure a magnitude-limited mass-period distribution. 12 of these systems are fully solved for the first time. A long-standing bias in the full population has been evident, favouring systems with orbital periods up to a few hours, with little exploration of the majority of the compact double white dwarf population, whose orbital period distribution centres at approximately 20\,hr. The 15 systems in this study span the orbital period range $5<P_\textrm{orb}<75$\,hr, significantly augmenting the number of well-characterised systems over these periods, and in general have two similar mass stars combining to $\approx$1.0\,M$_\odot$. We witness that the orbitally derived mass ratios generally show an excellent agreement with those deduced from atmospheric fits to double-lined spectra in previous work, emphasising the power of wide-scale spectroscopic surveys to efficiently locate the highest mass, double-lined double white dwarfs in the local Galaxy.

\end{abstract}




\section{Introduction}
Double white dwarf (DWD) binaries are one of the end-points of binary star evolution with roughly 100 million DWDs predicted to exist in the Milky Way alone \citep[][]{Marsh2011}. Two stars can evolve through a near isolated evolution in a wide binary \citep[][]{Heintz2024}, but also there exist many compact DWDs with orbital periods ranging from minutes to days. To be in a compact configuration, such systems have survived at least one common envelope phase \citep[][]{Paczynski1976, Webbink1984DWDprogenitorsRCrB}, making DWDs excellent probes of binary evolution as survivors of prior mass transfer. Furthermore, gravitational wave radiation causes the orbital separation to shrink over long timescales, making DWDs detectable with space-based, milli-hertz regime gravitational wave detectors \citep[e.g.][]{1987A&A...176L...1L, Nelemans2001gravitationalWaveSignal, Ruiter2010} such as the Laser Interferometer Space Antenna \citep[LISA,][]{LISA, LISAwhitepaper2023} or TianQin \citep[][]{TianQinProposal2016}. As a result, DWDs can give great insight to general relativistic, or non-general relativistic, orbital decay \citep[][]{Piro2019, Carvalho2022, Scherbak2023}. Another important characteristic of DWDs is their relevance as one of the long suspected progenitors of type \Romannum{1}a supernovae \citep[for a recent review, see][]{Ruiter2025}; enrichers of the galaxy through the production of heavy elements and are used as standard candles. Yet, largely owing to the difficulty in observing dim systems, full orbital solutions of very few DWD binaries have been obtained, and the constrained sample is observationally biased with inhomogenous selection criteria that are difficult to include in model comparisons \citep[e.g.][]{Li2019formationOfELMs}.

The current observational sample is far from reflective of the full mass-period distribution of DWD binaries, especially for compact binaries with an orbital period greater than a couple of hours, which is the overwhelming majority of DWDs \citep[e.g.][]{Nelemans2001closeWDs}. Worryingly, DWDs with total masses above 1.0\,M$_\odot$ are lacking in the observed sample, and with this the observed population has fallen short in providing type Ia progenitors that near the Chandrasekhar mass limit \citep[][]{Toonen2017, RebassaMansergas2019}, bringing into doubt whether the double-degenerate pathway is responsible for the majority of type Ia detonations \citep[e.g.][and references therein]{Liu2023}. The extent to which observational biases --- particularly those that favour the detection of DWDs with a low mass component --- play a factor in the lack of type Ia progenitors has long been the main impediment to a fully reflective picture.

The double-lined double white dwarf (DBL) survey was introduced in \citet{Munday2024DBL} and a crucial aspect of the survey is to handle observational biases in the compact DWD population by characterising a magnitude-limited sample with well-defined and understood selection criteria. Without such a deliberate effort, biases are difficult to handle, hampering the accuracy of comparisons to synthetic population models and making it difficult to draw conclusions about the full population of DWDs. As part of this survey, we obtained identification spectra and fitted atmospheric parameters to all observed star systems, resulting in at least 73\% of the 117 targets being DWD binaries of any kind. Specifically, the focus of the DBL survey is double-lined DWDs, being where a spectral signature from both stars is evident, for which at least a 29\% success rate of a source being double lined was retained. An unveiled class of previously rare massive total mass DWD binaries was recovered, including multiple near- or super-Chandrasekhar total mass DWDs \citep[][]{Munday2024DBL, Munday2025nature}. When brought into contact, the fate of the majority of detected systems is to initiate unstable mass transfer, resulting in the merger or pre-merger explosion of sources. The detection of double-lined sources is sensitive to mass ratios of $0.5\lessapprox q\lessapprox2$, with limits due to the fact that the dimmer star must contribute more than approximately 25\% of the flux to be spectroscopically disentaglable. 

Identification spectra of a large number of DWD candidates was the objective of this piloting work. What remained to be done was phase-resolved spectroscopy of the systems to obtain the orbital periods and hence the underlying period distribution of the double-lined DWD binaries in the sample. Many of the orbital periods are now presented in this study. The follow-up observations that we have undertaken are supplied in Section~\ref{sec:Observations}, our fitting methods in extracting RVs and orbital solutions in Section~\ref{sec:Methods}, and in Section~\ref{sec:Results} the results of our work.


\section{Observations}
\label{sec:Observations}
The purpose of our observations was to obtain phase-resolved spectra of the double-lined DA+DA DWDs discovered in \citet{Munday2024DBL}, where Balmer absorption lines of each star are spectrally disentanglable. As the orbital periods were largely unconstrained, having only the knowledge of a maximum orbital period, our observing strategy was to first obtain two series of 3 or 4 consecutive spectra with a gap of a couple of hours between them. Then, a third and fourth series of spectra would be taken in two of the following three nights to improve sensitivity to wider orbits. Together, this meant that we had a starting point of 12 to 15 spectra per target to resolve the orbital period. For systems where viable aliases were obtained in this time but the exact period alias was unclear, we obtained further observations at a later time with a targetted and dynamic strategy. For example, if strong period aliases appear at 10\,hr and 20\,hr, we would attempt to obtain spectra of the target with a gap of 5\,hr. Stronger evidence of the true orbital period would then be ascertained due to the different orbital phase coverages.

Time-series spectra were obtained across 22 nights using the Intermediate Dispersion Spectrograph (IDS) on the 2.5\,m Isaac Newton Telescope (INT). We used the R1200R grating centred on H$\alpha$ with a 1.0\arcsec slit, covering the wavelength range of 6\,000--7\,130\,\AA~with a spectral resolution of $R=6310$ at the centre of H$\alpha$. Observations were carried out over the nights 27 August 2019 -- 7 September 2019, 23 \& 24 September 2019, 7 June 2020 -- 14 June 2020. An arc frame was taken after acquisition of each target. For an extended duration on target, arcs were taken every 45\,min to correct for flexure. Later, after analysing the majority of these data, we obtained 9 nights on the INT from 25 April to 03 May 2024 to resolve the orbits of a few more systems and improve the solution of certain systems that had multiple viable period aliases. The first 4 nights utilised the H1800V grating with a 1.4$^{\prime\prime}$ slit width (with a DIMM seeing of about 1.0$^{\prime\prime}$), giving a spectral resolution of $R=9400$, and the other 5 used the same R1200R grating setup mentioned before. Additional spectra were obtained with the INT across various nights in the years 2023 and 2024 using the R1200B grating, providing a spectral resolution of about $R=4700$ at H$\beta$. These were reduced with the same method as the R1200R setup.

Bias frames and tungsten lamp flats were taken at the beginning of each night and used in the data reduction. Spectrophotometric standard stars were also observed at the start or the end of each night to flux calibrate the science exposures and correct for the instrumental response function. When multiple arc frames were taken of a target, the solution to the arcs was interpolated for the mid-exposure time of the science frame for wavelength calibration. All data were reduced using the \textsc{Molly} package \citep{Marsh2019Molly} and spectra were extracted using the method outlined in \citet{Marsh1989optimalExtraction}. Lastly, we utilise the spectra obtained from the William Herschel Telescope (WHT) of each target as was presented in \citet{Munday2024DBL}.

Once all of these data had been fully and thoroughly analysed, we decided to target orbital phases of three systems which were close to being solved using the Nordic Optical Telescope (NOT) across the nights 23--28 January 2025 and 21--25 May 2025. The employed instrument setup was ALFOSC with a g17 grating and a 0.75\arcsec slit width, giving wavelength coverage of H$\alpha$ at a spectral resolution of $R=7500$. An arc was taken before each exposure for wavelength calibration, and all data were reduced with the \textsc{pypeit} python package \citep[][]{pypeit}.

Finally, for the target WDJ231404.30+552814.11, nine spectra were taken with the Gemini Multi-Object Spectrograph (GMOS) on Gemini North as part of the queue program GN-2022B-Q-103. Observations were taken over the period 3-8 August 2022 with the R831 grating and a 0.5$^{\prime\prime}$ slit to give a spectral resolution of $R=4396$. 
Each spectrum had an exposure time of 8 minutes. A CuAr lamp spectrum was taken after each exposure for wavelength calibration. We used the IRAF GMOS package to reduce these data.

\section{Methods}
\label{sec:Methods}
\subsection{Synthetic spectra}
To begin with, we used the atmospheric solution of both stars derived in \citet{Munday2024DBL} and maintained the solution fixed to create a synthetic spectrum of both stars. As a base model, we utilised the 3-dimensional (3D) non-local thermal equilibrium (NLTE) grid first introduced in \citet{Munday2024DBL}, which uses the 3D grids of \citet{Tremblay2013, Tremblay2015} together with an NLTE correction that was presented in \citet{Kilic2021HiddenInPlainSight}. The NLTE correction was shown to significantly improve the fit to higher resolution data at H$\alpha$.

The synthetic and observed spectra were normalised to facilitate the processing of ground-based flux-calibrated data. In general, when T$_{\textrm{eff}} \approx 8\,000$\,K, we chose to normalise Balmer lines using data between wavelengths that correspond to velocities of $\pm(4\,500-5\,000)$\,km\,s$^{-1}$ for H$\alpha$, H$\beta$, H$\gamma$ and H$\delta$, and $\pm(2\,000-2\,500)$\,km\,s$^{-1}$ for H$\epsilon$ and H8. In WDs cooler than 8\,000\,K, we halved the normalisation range to be $\approx \pm(2\,250-2\,500)$\,km\,s$^{-1}$ and $\pm(1\,750-2\,000)$\,km\,s$^{-1}$, for the respective groups of Balmer lines.

\subsection{Improved atmospheric constraints on two systems}
\label{subsec:atmosphericWDJ1701}
Since publishing the first DBL survey instalment, we have obtained higher quality data of two systems presented in \citet{Munday2024DBL}. The first is WDJ181058.67+311940.94, which at the time of writing has the largest total mass of any DWD binary ($1.555\pm0.044$\,M$_\odot$). A devoted investigation of the system was presented in \citet[][]{Munday2025nature}. All data of WDJ181058.67+311940.94 obtained as part of the DBL follow-up were presented in that study, hence there are no updates to the atmospheric or orbital solution system here. For completeness, the orbital parameters of the system are also presented in this study, as it is a member of the double-lined systems detected in the piloting campaign.

The second system with improved data quality is WDJ170120.99$-$191527.57. Early analysis of the RVs indicated an orbital period of a few hours, flagging the binary as one of particular interest due to the fast RV variability, hence the extra observation. We used the Magellan Echellette \citep[MagE;][]{marshall2008-mage} spectrograph on the 6.5\,m Magellan Baade telescope at the Las Campanas Observatory to obtain two consecutive, 1200\,s long, high signal-to-noise ratio (SNR) spectra at an orbital phase where the system was at quadrature. The observations were performed using the 0.85$^{\prime\prime}$ slit, providing a wavelength coverage of 3700--9300\AA~at a resolving power of $R=4800$. We obtained three ThAr arc exposures immediately after our science exposures to generate an accurate wavelength solution. The spectroscopic data were reduced using \texttt{pypeit}. We also made use of the \texttt{merlin}\footnote{https://github.com/vedantchandra/merlin/tree/main} package which provides an end-to-end reduction pipeline based on \texttt{pypeit} v1.15.0 for MagE spectroscopic data. We hybrid fit these two spectra using an identical method to that presented in \citet{Munday2024DBL} with the 3D-NLTE synthetic spectra grid and the spectral fit is supplied in Appendix~\ref{fig:wdj1701_fit}. This indicates atmospheric parameters of T$_1=19780\pm270$\,K, $\log$(g$_1)=8.09\pm0.05$\,dex, T$_2=15490\pm260$\,K, $\log$(g$_2)=7.87\pm0.05$\,dex, which should be assumed in any continued analysis of the system. The quoted values include an external error of 1.4\% for T$_{\text{eff}}$ and 0.042\,dex for $\log g$ \citep[][]{Liebert2005} added in quadrature to the formal errors. The new star masses become M$_1=0.673\pm0.024$\,M$_\odot$ and M$_2=0.544\pm0.022$\,M$_\odot$, resulting in a slightly higher total mass ($1.217\pm0.032$\,M\,$_\odot$) than previously reported.

\subsection{Radial velocities}
To determine radial velocities (RVs), we take the best-fit atmospheric solution and keep it fixed, convolve the model to the resolution of the observation, and perform a least-squares minimisation with the data in the range of $\pm20\,\text{\AA}$ of H$\alpha$. We chose to isolate this wavelength region so that the goodness of fit to the wings of H$\alpha$ does not dominate, as the line core is the primary area of interest to fit RVs. Also, we fit to H$\alpha$ alone because of the much higher resolution in this spectral range for our observations and because the errors induced from imperfect wavelength calibration of data from the blue and red arms of ISIS exceed the benefits of its inclusion.

\begin{figure}
    \centering
    \includegraphics[width=0.49\columnwidth,clip,trim={0.65cm 0.6cm 0.65cm 0.65cm}]{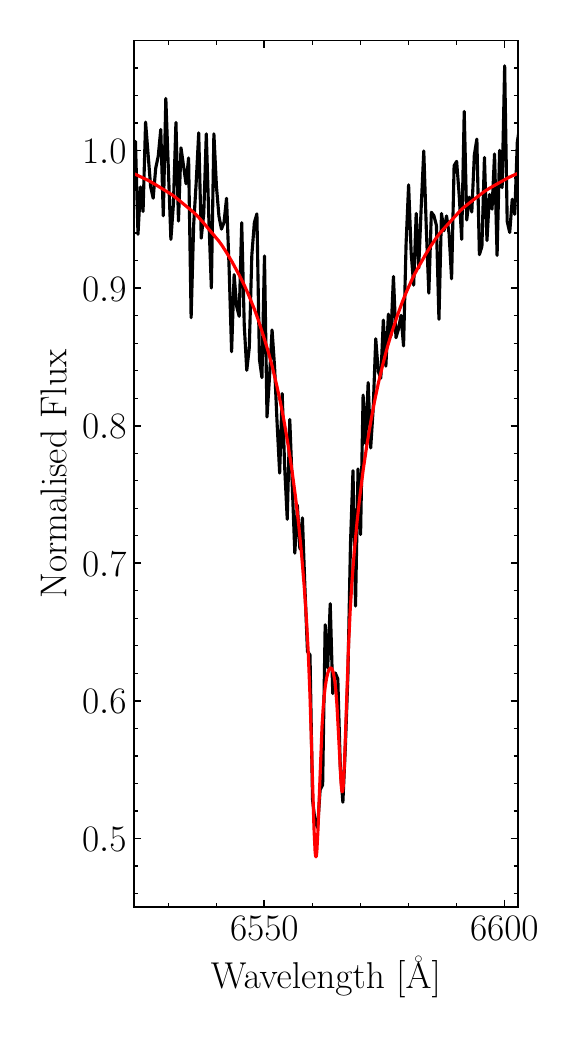}
    \includegraphics[width=0.49\columnwidth,clip,trim={0.65cm 0.6cm 0.65cm 0.65cm}]{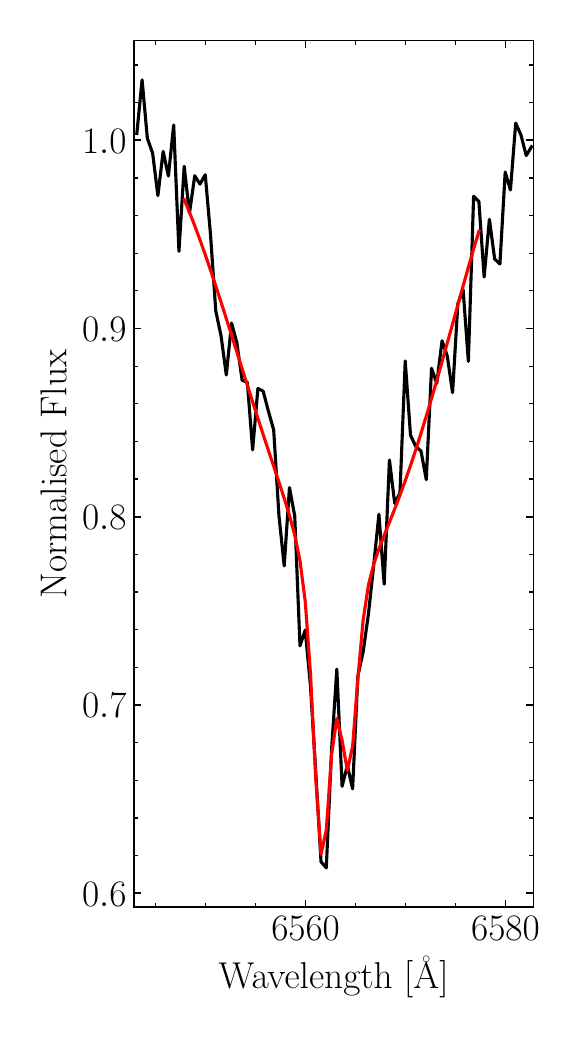}
    \caption{Example fits used to obtain RVs. The spectra are from the INT/IDS, reflecting the quality of the bulk of our data. \textit{Left:} A best-fit synthetic spectrum of WDJ005413.14+415613.73 with our formerly presented solution. Both stars have a similar line-core signature and the statistical RV errors for the blueshifted and redshifted stars were 4.5\,km\,s$^{-1}$ and 6.1\,km\,s$^{-1}$, respectively. \textit{Right:} A case for WDJ163441.85+173634.09 where fitting with two 2-Gaussian components plus an underlying polynomial resulted in a better template to fit to the observations. The statistical RV errors of the blueshifted and redshifted stars were 4.6\,km\,s$^{-1}$ and 7.1\,km\,s$^{-1}$, respectively. Generally, this approach slightly better fits the data but at the expense of losing spectral coverage, which becomes more important to utilise for lower ($\approx10-15$ in continuum) signal-to-noise observations. The normalisation procedure in the two cases is different, as described in the text.}
    \label{fig:RVfits}
\end{figure}

In a few cases (which are noted and commented on in Section~\ref{sec:Results}), we found that the template spectrum used for RV extraction improves when we instead fit 2-Gaussian profiles to the line cores of each star (that is, two 2-Gaussian profiles, one pair for each star) at the centre of H$\alpha$, on top of a fourth-order polynomial to model the shape of the broad absorption line within 20\,\AA~of its centre. This approach (or similar) has been frequently taken \citep[e.g.][]{Napiwotzki2002, Nelemans2005} as a 2-Gaussian model well approximates the shape of the centre of the Balmer absorption features, but the negative is that a fitted polynomial contains no physical information. We decided to employ this method for cases where the shape and depth of the synthetic line core profiles over- or under-fit the observed spectra across all instrumental setups. Doing so, we obtained a set of Gaussian parameters that are consistent to all spectra but smeared to the relevant spectral resolution\footnote{The reason for an imperfect atmospheric solution may lie with limitations in the input physics of synthetic grids, an inappropriately chosen mass-temperature-radius relation (carbon-oxygen core versus helium-core for WD masses $\approx0.4$--0.6\,M$_\odot$), which would compromise the apparent-to-absolute-flux scaling factor when fitting photometric data, or a faint tertiary component contributing a small percentage of flux, all of which leading to a too deep/shallow synthetic line core signature}. A comparison of fits with synthetic spectra and the 2-Gaussian approach is plotted in Fig.~\ref{fig:RVfits}.

Often, and in our INT data especially, when the site conditions worsened, the continuum SNR is relatively low ($\text{SNR}<20$), leading to an imperfect spectrum normalisation. To optimise the normalisation, we minimise the residual between the data and the atmospheric fit within $20\,\text{\AA}$~of the centre of H$\alpha$. Data $\pm5-20\,\text{\AA}$ away from the centre of H$\alpha$ was sigma-clipped with a threshold of 4$\sigma$; no data within $5\text{\AA}$ of the centre was clipped. RVs and RV errors are reported from the median and standard deviation of 1000 bootstrapping iterations. To give an idea of the effect of this optimisation, we take the double-lined system WDJ114446.16+364151.13 that has spectra with a wide SNR range and a near-identical spectral signature for both stars at H$\alpha$. For a noisy spectrum that has $\text{SNR}=15$ in the continuum and $\text{SNR}=8$ at the line core centre, the RV errors with no optimising of the normalisation are $\pm29.6$\,km\,s$^{-1}$ and $\pm20.1$\,km\,s$^{-1}$, and with optimisation give errors of $\pm22.6$\,km\,s$^{-1}$ and $17.7$\,km\,s$^{-1}$ for the two stars. Now, taking the same system but looking at data with $\text{SNR}=40$ in the continuum and $\text{SNR}=25$ at the line core centre, no optimising gives $\pm4.8$\,km\,s$^{-1}$ for both stars, while optimising gives $\pm5.0$\,km\,s$^{-1}$ for both stars. In the low SNR data case, the reported RV errors become appreciably better with optimisation, whereas the errors are nearly identical when the SNR is higher. Hence, applying this optimised normalisation method slightly improved the accuracy and precision of all RVs from low SNR data but made little difference to high SNR data. We furthermore note that the scatter in the residual of a sinusoidal fit to the RV curves (which is addressed in Section~\ref{subsec:OrbitalSolutions}) decreased for an optimised normalisation, indicating an improved accuracy in the extracted RVs.

\subsection{Orbital solutions}
\label{subsec:OrbitalSolutions}
A couple of the candidate double-lined systems with time-series spectroscopy are indeed double-lined, but others turned out to be single-lined DWDs. In searching for single-lined RV variability, we invoke the same procedure as in \citet{Munday2024DBL}. The mean of all extracted RVs for a source is taken and a null-hypothesis that the RV is a constant with respect to the mean is tested. We compute the $\chi^2$ of all measurements compared to the mean and use the relevant $\chi^2$-distribution for the number of degrees of freedom to calculate the probability that a source is not RV variable, setting a 4$\sigma$ outlier threshold for variability, equivalent to a false positive probability of 0.62\%.

For all RV variable sources, double- or single-lined, we searched for a periodic signal when more than 5 spectra were taken. This was performed with a least-squares minimisation with trial orbital frequencies, $f$, spaced in a way that no more than 0.01~cycles are skipped over the full baseline of data. The $\chi^2$ of each trial minimisation was analysed for a best-fit solution and the RVs (viewed at time $T_{RV}$) were modelled following
\begin{equation}
    V_{1,2}(\phi)=\gamma_{1,2} \pm K_{1,2}\sin{2\pi\phi}
    \label{eqn:Velocity_fn_phase}
\end{equation}
\noindent where $\phi=\left((T_{RV}-T_0)~\mathrm{mod} f\right)$ is the orbital phase for each epoch, $K$ is the semi-amplitude and $\gamma$ is the RV offset for each star denoted by subscripts 1 and 2. When analysing periodograms, we normalise the $\chi^2$ of the best trial solution for each frequency to obtain a power spectrum following \citep{Cumming2004}:
\begin{equation}
    z_{\textrm{Kep}}(f) = \frac{N - 5}{4} \frac{\chi_0^2 - \chi^2\left(f\right)}{\chi^2\left(f\right)}
    \label{eqn:ZkepCummingPowerNorm}
\end{equation}
with $N$ the number of epochs, $\chi_0^2$ the $\chi^2$ with respect to the weighted mean of the RV measurements (whereas $\chi^2$ is with the best Keplerian orbital fit). For double-lined fits, the RVs and weighted means from the two stars are handled separately and summed after. When the power $z_{\textrm{Kep}}$ is maximised, the $\chi^2$ is minimised.

A strength of having the unique spectral signature and atmospheric parameters of both stars in double lined sources is that we can furthermore use physical insight of the system to limit the maximum constraint of $K_{1,2}$ for each trial frequency. A minimum and maximum $K_{\textrm{max}}$ can be inferred from the binary mass function with the expectation that the stars are circularised, using
\begin{equation}
    P K_{\textrm{max}}^3 = \frac{2\pi G\textrm{M}_{\textrm{max}}^3 \sin(i=90^\circ)^3}{(\textrm{M}_{\textrm{min}} + \textrm{M}_{\textrm{max}})^2}
    \label{eqn:BinMassFrac}
\end{equation}
\noindent where, for a DWD system, a sub-Chandrasekhar mass limit M$_{\textrm{max}}=1.4$\,M$_\odot$ and a minimum WD mass of M$_{\textrm{min}}\approx0.15\,$M$_\odot$ \citep{Istrate2016, Calcaferro2018coolestELMtracks} applies to put upper and lower limits on $K_{1,2}$, respectively. For single-lined WD binaries, the companion may be a neutron star with maximum mass of approximately 3.0\,M$_\odot$ \citep[e.g.][]{1996ApJ...470L..61K}, hence M$_{\textrm{min}}=0.15\,$M$_\odot$ and M$_{\textrm{max}}=3.0$\,M$_\odot$ applies. Black hole companions to a WD are not expected in the sample due to their much smaller predicted number in the Galaxy \citep[][]{Nelemans2001gravitationalWaveSignal} and would only be applicable to systems that spectrally appear as a single WD (no excess flux). Additional restrictions can be applied to place limits on the orbital inclination, $i$, if photometric variability is or is not witnessed (Section~\ref{subsec:ligtcurves}), causing $PK_{\textrm{max}}^3$ to alter. Similarly in this case, a constraint on $PK_{\textrm{min}}^3$ could be applied.

We also limited the range of valid orbital solutions based on the details of the spectral fit, again possible from the double-lined nature of these sources. The difference between $\gamma_1$ and $\gamma_2$ is equal to the difference in gravitational redshifts and, as such,
\begin{equation}
    \gamma_2 - \gamma_1 = \frac{G}{c}  \left(  \frac{M_2}{R_2}  -  \frac{M_1}{R_1} \right) =  \frac{G}{c}  \left(    \frac{K_1 M_1/K_2}{R_2}  -  \frac{M_1}{R_1}  \right)
\end{equation}
\noindent with star radii ($R$) predictable from mass-radius relationships. With this and the atmospheric solutions for each system, we decided to take the conservative approach of applying a flat prior on the relative gravitational redshift spanning 3$\sigma$ from the median solution for both the masses and the radii of each star. The typical mass errors are $\pm$0.04\,M$_\odot$ and, while we before assumed a thick hydrogen envelope consistent with single star evolutionary tracks to scale to an Eddington flux, the size of the hydrogen envelopes is largely unknown. With that in mind, 3$\sigma$ errors on the interpolated radii were again appropriate to incorporate the uncertainty in the model fit. We note that this is an intentionally cautious approach and was chosen in case that an incorrect atmospheric solution is found from a local minimum. The restriction on the difference of gravitational redshifts served to remove spurious periodogram solutions when finding a best-fit solution.

\begin{figure}
    \centering
    \includegraphics[width=\columnwidth,clip,trim={0.6cm 0.75cm 0.5cm 0.5cm}]{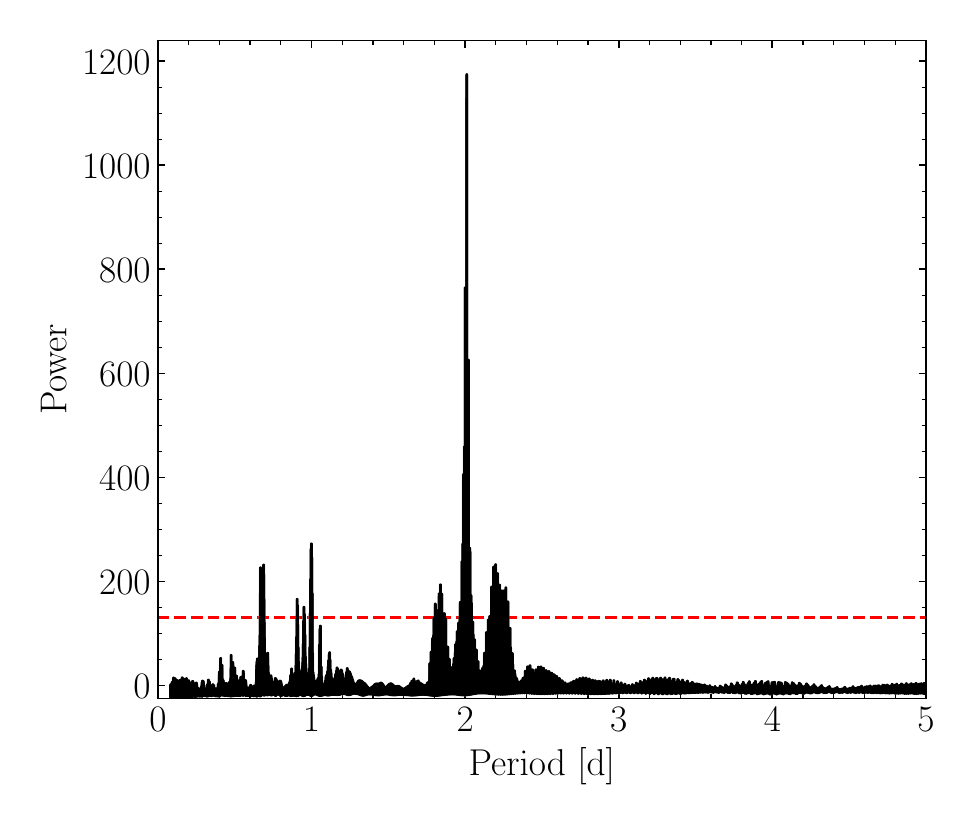}
    \caption{An example Lomb-Scargle periodogram showing that of WDJ020847.22+251409.97. The spacing between trial frequencies is chosen such that no more than 0.01~cycles are skipped across the full time baseline between adjacent solutions. The dashed red horizontal line shows the false alarm probability at a 4$\sigma$ level. Multiple peaks lie above the 4$\sigma$ level and are a consequence of the wide spacing of observation and certain aliases that better fit the RVs of just one of the stars. However, the non-highest peaks clearly show blocks of observations that lie off the sinusoidal trend upon visual inspection. This isolates the true orbital period of 2.008\,d (48.1837\,hr).}
    \label{fig:ExamplePeriodogram}
\end{figure}

We calculated false alarm probabilities for each system to analyse the significance of each peak in the power spectrum. This was performed through a bootstrap approach by randomly shuffling the timestamps of observation, re-calculating Lomb-Scargle periodograms \citep[][]{Lomb1976,Scargle1982} over 50 iterations and taking quantiles of the full array of powers from these 50 iterations as the false alarm probability. We consider a trial period to be the true period of the system if there is a unique peak above a 4$\sigma$ false alarm probability and 3$\sigma$ for a peak to be considered an alias or the true period. Below those thresholds, the period is not completely solved but we assign limits on the maximum/minimum orbital period based on the information currently extractable by taking the upper/lower boundaries where no period aliases fall above a 2$\sigma$ false alarm probability. Here, further RV measurements would be required to fully solve the orbits. In cases where there is more than one peak above the 4$\sigma$ or 3$\sigma$ alarm probability, we inspect all individual peaks to look for a cause and assess if any of these solutions are suitable. When this happened, it often was an artifact of the wide time spacing between observations, but all cases where where this occurred are commented on in Section~\ref{sec:Results}. An example periodogram for a solved system is found in Fig.~\ref{fig:ExamplePeriodogram}.

\subsection{Light curves}
\label{subsec:ligtcurves}
We analysed time-series archival photometry from the Transiting Exoplanet Survey Satellite \citep[TESS,][]{TESS}, Zwicky Transient Facility \citep[ZTF,][]{ZTF}, Catalina Real-time Transient Survey \citep{CRTS}, \textit{Gaia} \citep{GaiaDR3_2023} and Asteroid Terrestrial-impact Last Alert System \citep{ATLAS} to search for photometric variability amongst the full selection of DWDs\footnote{https://github.com/JamesMunday98/AllSurveyPhotometry}. The brightest of our sources saturate in some of these surveys and, when they are saturating, time-series photometry was ignored. Sources in \textit{Gaia} listed as variable have available photometric data for download as of data release 3 and in these cases the \textit{Gaia} photometry was analysed. \textit{TESS} light curves in the fast, short and long cadences were obtained when available from the MAST system, and we extracted the photometry from \textit{TESS} pixel frames when not available (typically for sources dimmer than 16th mag).

Finally, we note that \textit{TESS} (all cadences) and \textit{ZTF} (in filters $g, r, i$) photometry was inspected for all systems with solved periods by searching for variability within 2\% of the orbital period. The same was performed for all aliases of systems that have more than one above the false alarm probability, described in Section~\ref{subsec:SystemsPeriodAliases}. In no case was a significant photometric signal observed for a 1-term Lomb-Scargle or a box least squares periodogram. Usage of a 1-term Lomb Scargle periodogram is most appropriate for identifying Doppler beaming, tidal deformation or irradiation, which will be the strongest orbitally-induced variability in all systems \citep[][]{Hermes2014}, whereas a box least squares periodogram could reveal the unlikely event of eclipses, requiring an almost perfectly edge-on inclination for the typical DWD binary. Gravitational lensing is unlikely to occur or be detected \citep[][]{Marsh2001lensing, Sajadian2025}.

\section{Results}
\label{sec:Results}
We present here an overview of all systems that now have full atmospheric \citep{Munday2024DBL} and orbital solutions. This section speaks on newly presented results as well as ones obtained in prior work  \footnote{In \citet{Munday2025nature}, we noted that the RVs of spectra with spectral resolution $R<10\,000$ were slightly inaccurate ($5-10$\,km\,s$^{-1}$) when spectral lines perfectly overlap at the time of observation and were removed when solving for the orbital solution. Here however, we choose not to disregard any RVs that fall into this category since, with approximately 15--20 RVs per system, the information gained by including them in settling period aliases is far beneficial. One should consider their removal if more RVs are obtained in future work.}. We particularly emphasise here that the full population of compact DWDs are expected to have a median period of about 20\,hr, as found through synthetic binary population synthesis models \citep[][]{Nelemans2001closeWDs}. 

It should be noted that inter-night observations cause a decreased efficiency in solving orbital periods because of the consequent window function. For example, periods of 8\,hr, 12\,hr, 24\,hr, 48\,hr, etcetera, are more challenging to solve. Multiple day orbital periods are also more challenging to ascertain because our strategy of taking 3 consecutive spectra does not cover as large a range of orbital phase. The selection of systems from the 34 double-lined systems we found in \citet{Munday2024DBL} is largely random, with the only restriction being the visibility for the scheduled telescope nights and the need for both stars to be separable at a spectral resolution of $R=6310$. Lastly, there is sensitivity to the full range of orbital periods above 2\,hr (with this minimum set because of orbital smearing across 30\,min exposure times) and orbital periods are sampled up to 10\,d.

\subsection{Solved periods}
\label{subsec:resultsSolvedPeriods}
\begin{figure*}
    \centering
    \includegraphics[width=0.45\textwidth, clip, trim={0cm 0cm 1cm 0.5cm}]{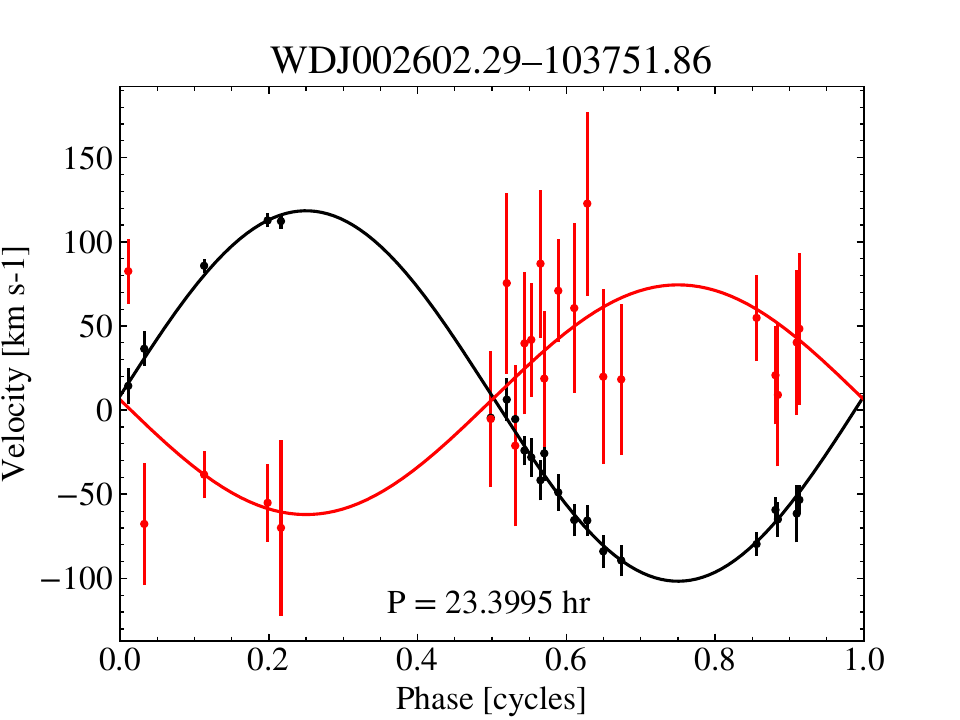}
    \includegraphics[width=0.45\textwidth, clip, trim={0cm 0cm 1cm 0.5cm}]{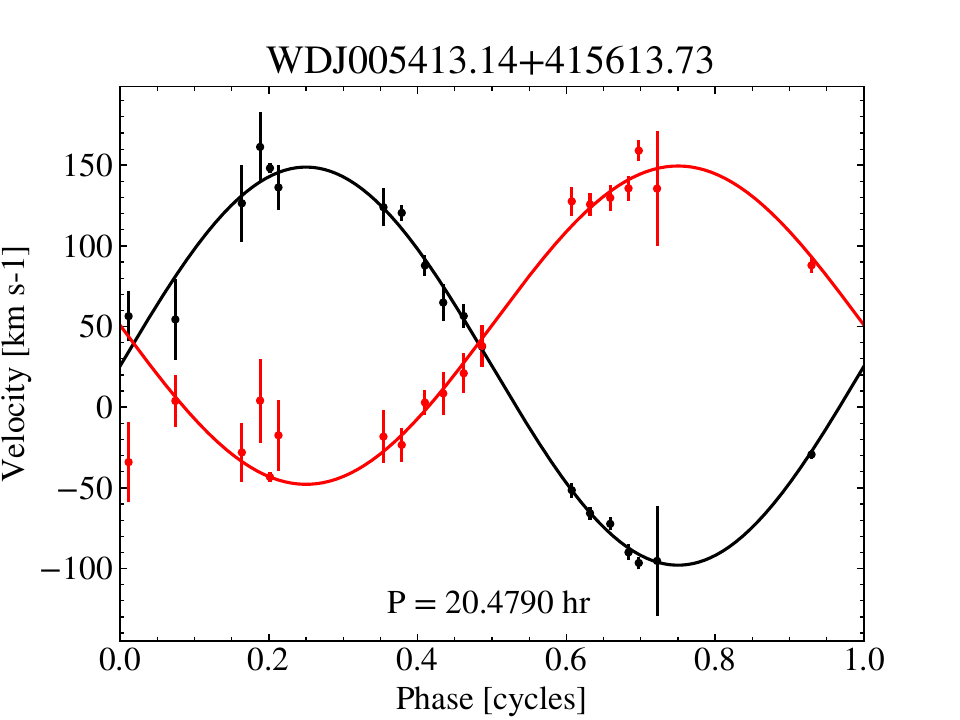}
    \includegraphics[width=0.45\textwidth, clip, trim={0cm 0cm 1cm 0.5cm}]{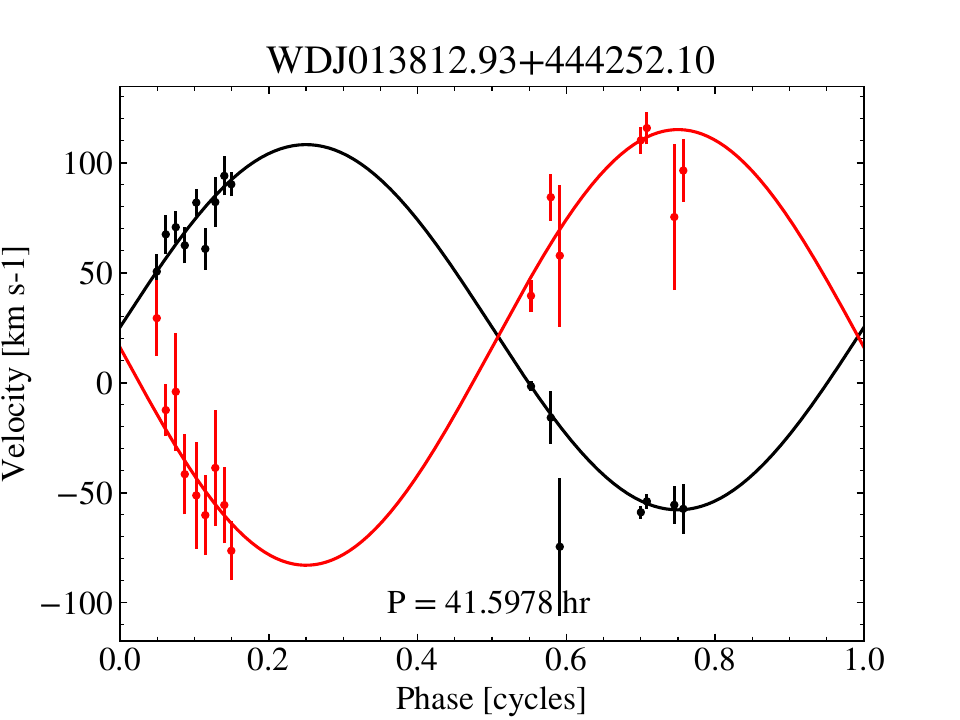}
    \includegraphics[width=0.45\textwidth, clip, trim={0cm 0cm 1cm 0.5cm}]{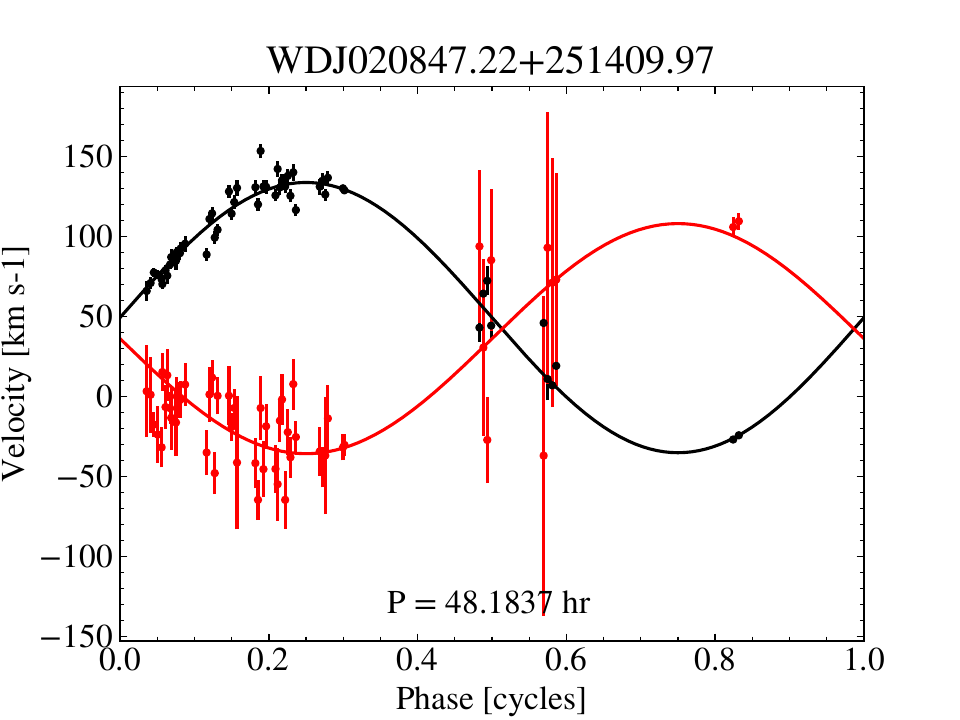}
    \includegraphics[width=0.45\textwidth, clip, trim={0cm 0cm 1cm 0.5cm}]{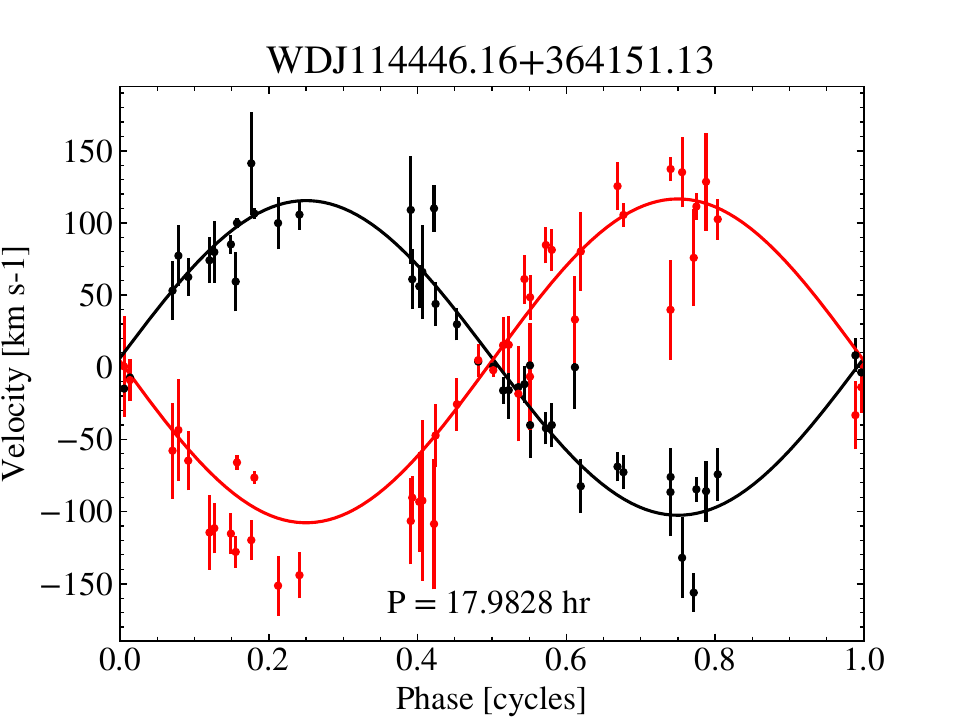}
    \includegraphics[width=0.45\textwidth, clip, trim={0cm 0cm 1cm 0.5cm}]{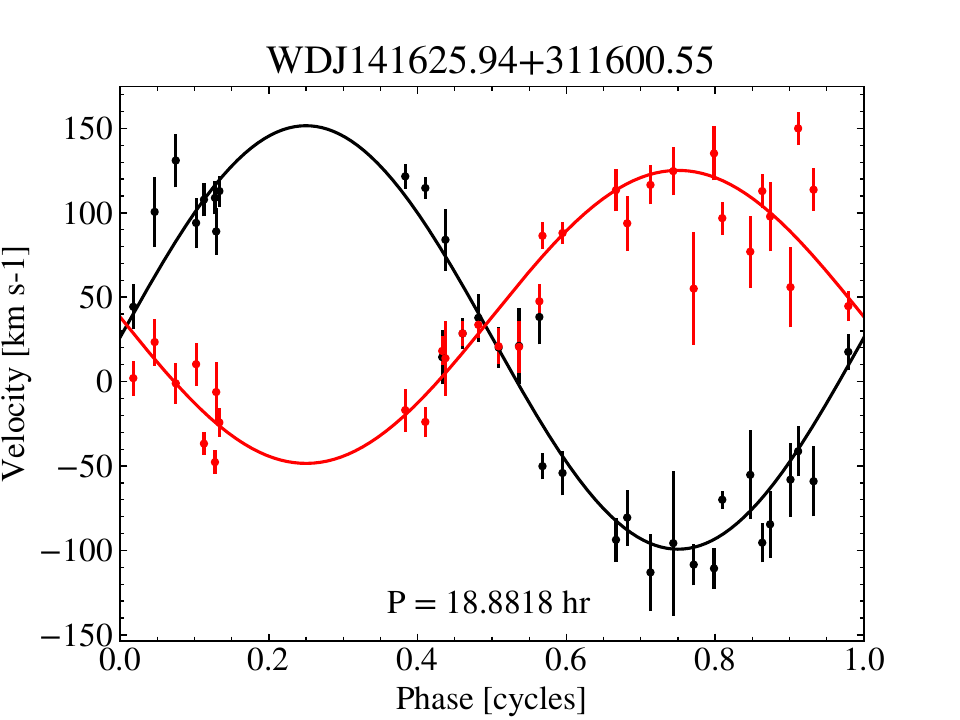}
    \caption{RV curves of all double-lined systems with orbital solutions presented in this study. The data points and the curves in black represent the hotter star and those in red are for the cooler star. The details of each solution is given and described in text. Titles above each plot show the system that each curve corresponds to, while the orbital period is mentioned at the bottom of each plot for convenience. The velocity for each point here has been corrected for relativistic effects and therefore is not an observed velocity (Appendix~\ref{appendix:RVs}).}
    \label{fig:AllRVcurves1}
\end{figure*}

\begin{figure*}
    \centering
    \includegraphics[width=0.45\textwidth, clip, trim={0cm 0cm 1cm 0.5cm}]{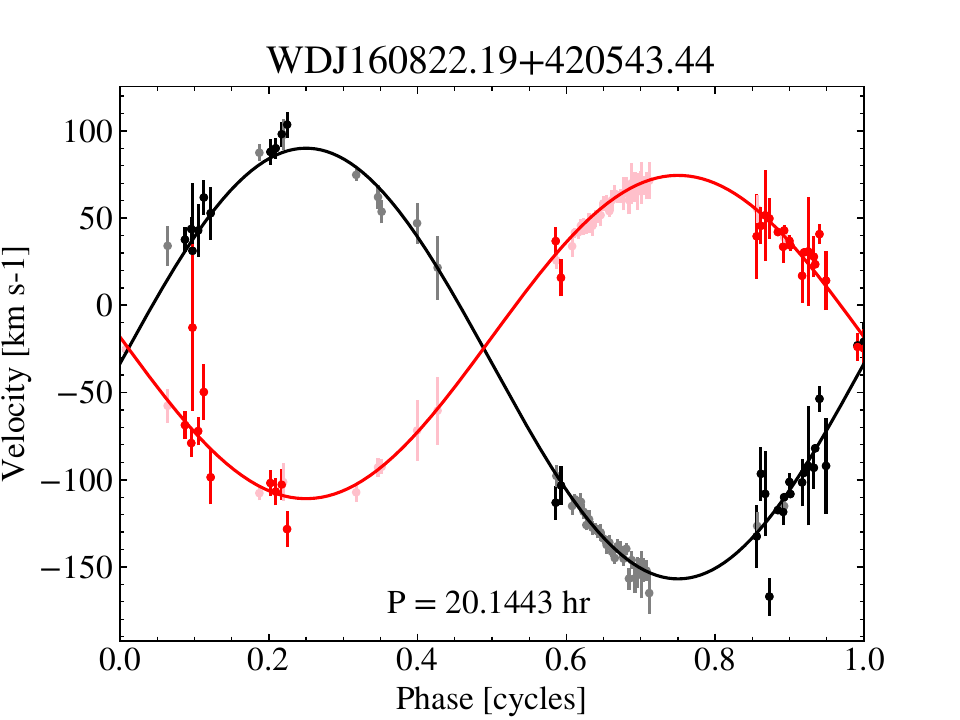}
    \includegraphics[width=0.45\textwidth, clip, trim={0cm 0cm 1cm 0.5cm}]{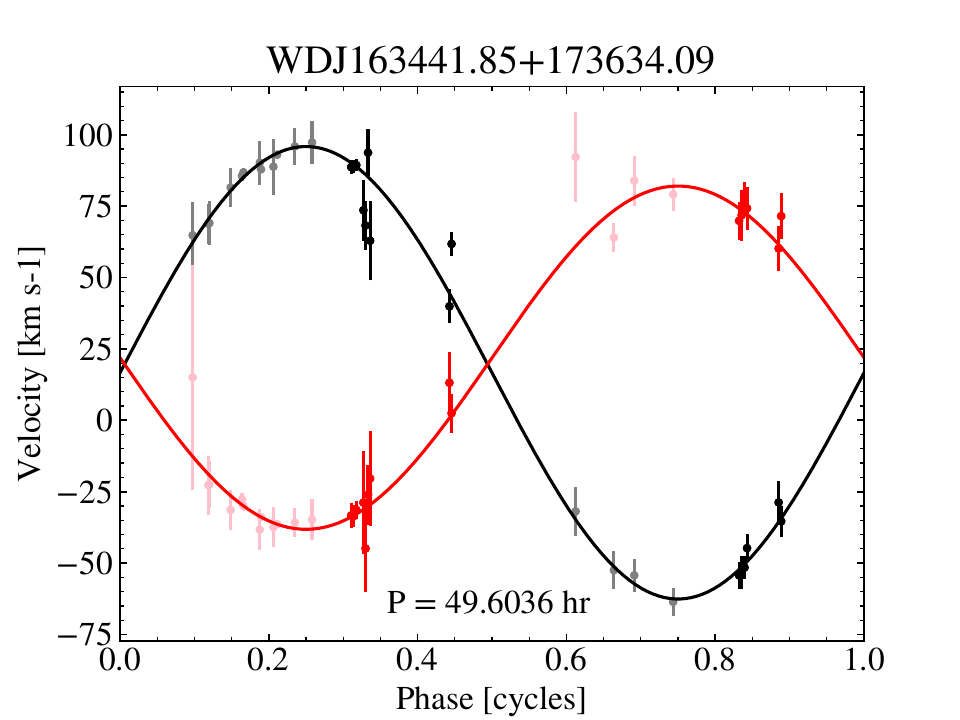}
    \includegraphics[width=0.45\textwidth, clip, trim={0cm 0cm 1cm 0.5cm}]{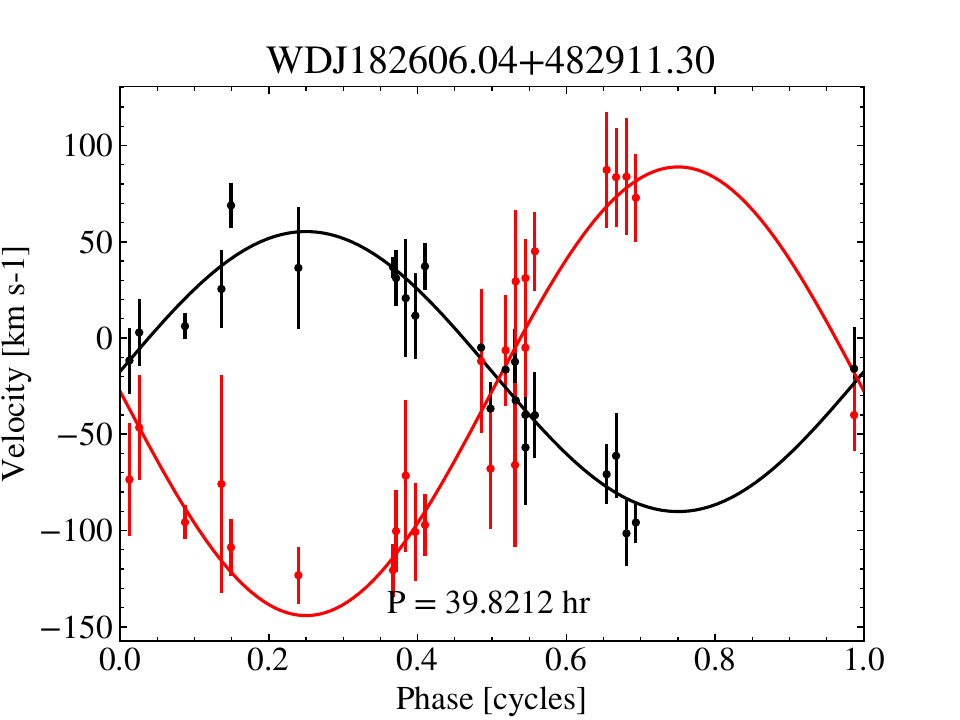}
    \includegraphics[width=0.45\textwidth, clip, trim={0cm 0cm 1cm 0.5cm}]{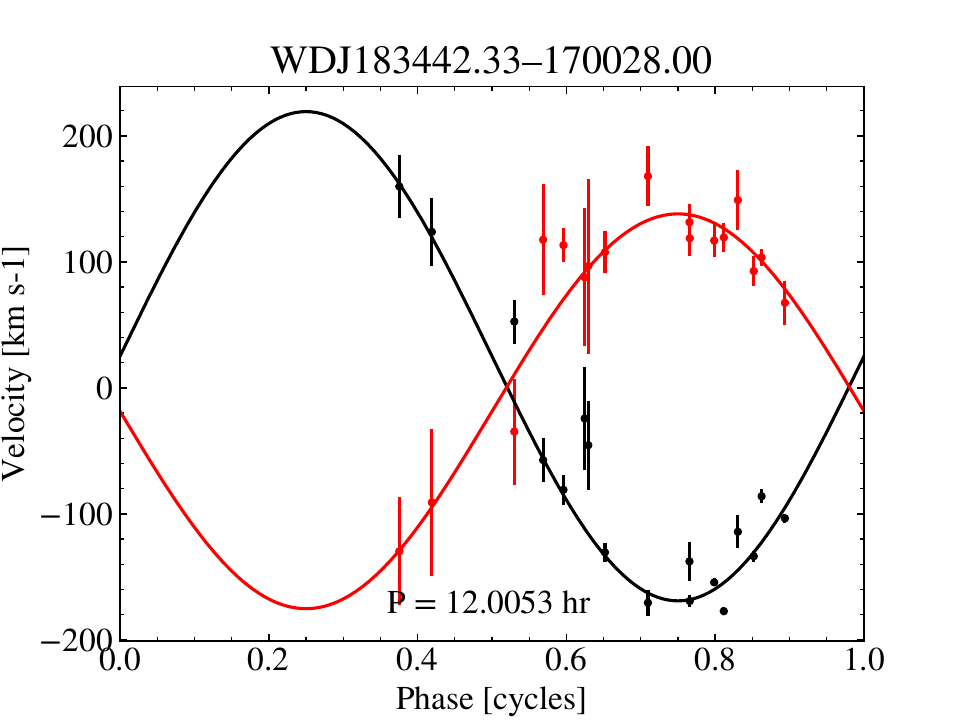}
    \includegraphics[width=0.45\textwidth, clip, trim={0cm 0cm 1cm 0.5cm}]{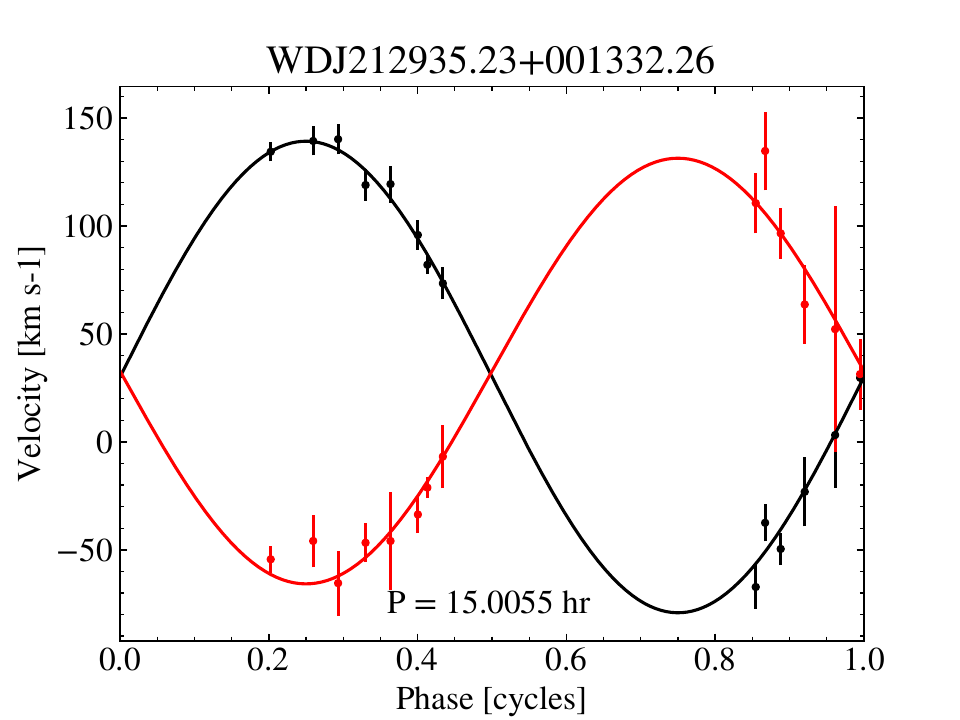}
    \includegraphics[width=0.45\textwidth, clip, trim={0cm 0cm 1cm 0.5cm}]{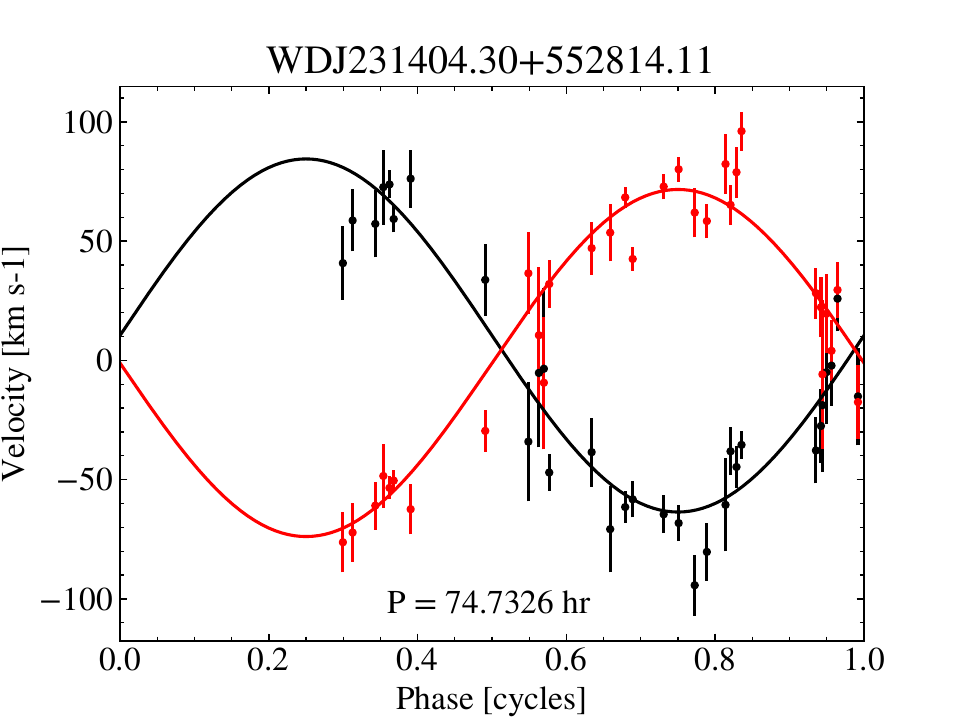}
    \caption{RV curves of the other six systems presented in this work, continuing from Fig.~\ref{fig:AllRVcurves1}. The data for the hotter star of the binary is plotted black, and the cooler star in red. WDJ153615.83+501350.98 is omitted from the plots as we only had two further RV measurements to add to those in \citet{Kilic2021HiddenInPlainSight}, whereas for WDJ160822.19+420543.44 and WDJ163441.85+173634.09 we were able to recover the same orbital solutions in this work with a much larger number of RVs to add. The RVs presented in \citet{Kilic2020twoDoubleLined} and \citet{Kilic2021HiddenInPlainSight} for these two systems were incorporated into our analysis, plotted in light red/black shades. The system WDJ151109.90+404801.18 is also omitted given its unclear orbital solution, even though the correct period alias has been identified.}
    \label{fig:AllRVcurves2}
\end{figure*}

The list of fully solved systems are as follows, with comments on the validity of the results when comparing the atmospheric and orbital solutions. The RV curves for each are plotted in Fig.~\ref{fig:AllRVcurves1} and Fig.~\ref{fig:AllRVcurves2}. Merger times are mentioned when the orbital period is below 15\,hr using Peters' equation \citep[][]{Peters1964GravRadTwoPointMasses} for a circular orbit. The double-lined systems here are 15 of the 34 presented in \citet[][]{Munday2024DBL}, while the first discovery of 3 of them was in other studies. RVs for each source are presented in Appendix~\ref{appendix:RVs}.

\textbf{WDJ002602.29$-$103751.86} has one peak above 4$\sigma$ and one other close to this threshold, lying at 23.40\,hr and 7.91\,hr respectively. The first of these clearly best fits the RVs of the brighter star and has a near-zero difference in gravitational redshift, while the other solution slightly better fits RVs for the dimmer star and has a gravitational redshift difference of 10--20\,km\,s$^{-1}$. We strongly expect that the true period is 23.40\,hr because of the better fit and accordance with the atmospheric solution, and hence adopt this value. This leads to the system having an orbitally derived mass ratio of $q=K_2/K_1=0.62\pm0.27$, while the atmospheric mass ratio is $q=M_1/M_2=1.12\pm0.07$, which is significantly different. This difference cannot be resolved by forcing a carbon-oxygen or helium core mass-radius relationship in the atmospheric fit, as this still leads to a poor match to the observations for all combinations. The detection of the dimmer star is marginal at H$\alpha$ and the fractional flux contributed in other wavelengths is small -- further RVs, and perhaps higher signal-to-noise observations at quadrature with H$\alpha$, are encouraged to improve the accuracy of $K_2$. The Gaussians plus polynomial template proved advantageous in obtaining more precise RVs due to the slight detection of the cooler WD in this system.

\textbf{WDJ005413.14+415613.73} has an orbital period of 20.48\,hr, with an orbitally-derived mass ratio $q=K_2/K_1=0.78\pm0.08$ and the atmospheric solution represents $q=0.96\pm0.12$.

\textbf{WDJ013812.93+444252.10} has an orbital period of 41.60\,hr, with an orbitally-derived mass ratio $q=K_2/K_1=1.19\pm0.27$ and the atmospheric solution represents $q=1.08\pm0.07$.

\textbf{WDJ020847.22+251409.97} is the closest new double-lined DWD in the DBL survey, located at 39\,pc, formerly identified as a double degenerate candidate in the 40\,pc survey of WDs \citep[][]{McCleery2020}. We find that it has an orbital period of 48.18\,hr, the atmospheric solution represents $q=M_1/M_2=0.74\pm0.04$ and the orbital solution $q=K_2/K_1=0.85\pm0.11$. The Gaussians plus polynomial method was used to extract RVs, again because of the slight detection of the cooler WD.

\textbf{WDJ114446.16+364151.13} has an orbital period of 17.98\,hr, with an orbitally-derived mass ratio $q=K_1/K_2=1.03\pm0.15$ and the atmospheric solution represents $q=0.93\pm0.09$.

\textbf{WDJ141625.94+311600.55} has an orbital period of 18.88\,hr with an orbitally-derived mass ratio $q=K_2/K_1=0.69\pm0.12$ and the mass ratio from the atmospheric solution is $q=1.23\pm0.10$. The two stars are near identical in terms of their atmospheric parameters and their spectral signature. If we switch the RV assignments such that the hotter star is given the RVs of the previously thought-to-be cooler star, we get atmospheric parameters of T$_{1,\textrm{eff}}=13\,460\pm400$\,K, T$_{2,\textrm{eff}}=12\,740\pm250$\,K, $\log(g_1)=7.66\pm0.16$\,dex, $\log(g_2)=7.70\pm0.15$\,dex, $M_1=0.44\pm0.05$\,M$_\odot$ and $M_2=0.45\pm0.05$$_\odot$. This generates an atmospheric mass ratio of $q=0.98\pm0.16$, which far better aligns with the orbital solution, hence we assume these new atmospheric parameters going forward.

\textbf{WDJ151109.90+404801.18} is an interesting case where, for 5 of the 23 spectra (3 on one night, 2 on another), we are not able to determine which star is which in the double-lined spectra even when trialling every valid combination of RV assignments. The reason for this is that the spectral signature from both stars at H$\alpha$ is very similar. However, what is common to all trial combinations is that an orbital period of approximately 23.3--23.6\,hr is clearly the highest and the only high in power peak, making an orbital period in this range the correct solution for this binary. The short-term RV variability of a group of six spectra on the same night covering a 4.8\,hr observing window affirms that the near 1 day orbital period is a true detection. At times, and depending on the assignment of the unclear RVs, another peak at half of this period emerges, but this well fits the RVs of only one of the stars. For the 23.3--23.6\,hr period aliases, we find that $q=K_2/K_1\approx1.0$--$1.5$ in all cases, but the error on $q$ is approximately $\pm0.4$--$0.5$ and so it is largely unconstrained. The high percentage error stems from the poor phase coverage given the near 1\,d orbital period, though, because of the similar spectral signature from both stars, conclusively resolving the orbital parameters of this binary would be very challenging unless devoting entire telescope nights to its observation. Noted in Appendix~\ref{appendix:RVs} are the RVs that are unclear. The atmospheric solution for this system gives $q=1.52\pm0.11$, which is in the general vicinity of the orbital solution. The Gaussians plus polynomial method was used to measure RVs since the template fit to the line core shapes was improved for these stars with a similar line core signature.

\textbf{WDJ153615.83+501350.98}, or \textbf{WD\,1534+503}, was discovered and published before completion of this work \citep[][]{Kilic2021HiddenInPlainSight}. We only have 2 additional RV measurements from the identification spectra for this system, which we obtained using the Gaussians plus polynomial method. Combining datasets, we find $P_{\textrm{orb}}=17.02$\,hr. This orbital period has a greatly improved precision in comparison to previously published work given the extra year in the time baseline of the new RV measurements. We next quote our orbital solution, though emphasise that differences in the fitting procedure between their work and ours, and the instrumental setups/data reduction, could lead to systematic errors in the results. We find $K_1=135.08\pm4.27$\,km\,s$^{-1}$, $K_2=89.15\pm9.88$\,km\,s$^{-1}$, $\gamma_1=23.22\pm3.22$\,km\,s$^{-1}$, $\gamma_2=44.45\pm7.00$\,km\,s$^{-1}$. The orbital solution hence indicates $q=K_2/K_1=0.660\pm0.076$. This remains in good agreement with the atmospheric solution presented in \citet{Kilic2021HiddenInPlainSight}, which was $q=M_1/M_2=0.64\pm0.15$, and less so with that of \citet{Munday2024DBL}, which was $q=M_1/M_2=0.96\pm0.07$. The difference likely stems from the choice of mass-radius relationships employed, in such a way that the lower mass star's radius does not align well with a helium-core WD model. It may be a carbon-oxygen or a hybrid core, and the orbital solution hence encourages adoption of the atmospheric solution in \citet[][]{Kilic2021HiddenInPlainSight}.

\textbf{WDJ160822.19+420543.44}, or \textbf{WD\,1606+422}, was independently discovered and published before completion of this work \citep[][]{Kilic2020twoDoubleLined}. To verify the orbital solution found by these authors with a unique dataset, we used our 30 RV measurements of the binary and found a solution consistent with their results. Maintaining the same convention that star 1 is the hotter star and given the excellent agreement, we decided to merge the datasets to improve the orbital solution further. Overall, we arrive at a solution of a 20.14\,hr orbital period with $K_1=92.60\pm2.45$\,km\,s$^{-1}$, $K_2=127.24\pm2.30$\,km\,s$^{-1}$, $\gamma_1=-18.04\pm1.65$\,km\,s$^{-1}$, $\gamma_2=-31.40\pm1.57$\,km\,s$^{-1}$. With this, the newfound mass ratio from the orbital solution is $q=K_2/K_1=1.37\pm0.04$. Our atmospheric solution suggests $q=M_1/M_2=1.17\pm0.08$, and with this all results are consistent.

\textbf{WDJ163441.85+173634.09}, or \textbf{PG 1632+177}, was also independently discovered and published by \citet{Kilic2021HiddenInPlainSight} and again we find consistency between our atmospheric solution and theirs. Given a non-ideal atmospheric fit to the H$\alpha$ line cores for this binary, we decided to fit the RVs with the Gaussians plus polynomial method. The orbital solution from the RVs from both studies with our additional 14 RV measurements gives an orbital period of 49.60\,hr, $K_1=79.63\pm1.73$\,km\,s$^{-1}$, $K_2=-60.59\pm1.74$\,km\,s$^{-1}$, $\gamma_1=16.21\pm1.79$\,km\,s$^{-1}$, $\gamma_2=22.38\pm1.44$\,km\,s$^{-1}$. This system is another case where the mass of one of the stars is on the boundary between a CO or He-core WD, resulting in a large discrepancy between $M_1/M_2$ and $K_2/K_1$. The solution presented in \citet[][]{Kilic2021HiddenInPlainSight} better fits the data for which the lower mass WD with $M_1=0.392^{+0.069}_{-0.059}$\,M$_\odot$ has a hybrid He/CO or a CO core and the larger mass WD has a mass $M_2=0.526^{+0.095}_{-0.082}$\,M$_\odot$. We continue to encourage usage of this result, making the atmospheric $q=M_1/M_2=0.75\pm0.18$ and the newly-found orbital with an improved precision on the orbital period $q=K_2/K_1=0.761\pm0.027$.

\textbf{WDJ181058.67+311940.94}, as presented in \citet{Munday2025nature}, has an orbital period of 14.24\,hr, with an orbitally-derived mass ratio $q=K_2/K_1=0.98\pm0.03$ and the updated atmospheric solution represents $q=0.86\pm0.04$. Its merger time is $22.6\pm1.0$\,Gyr.

\textbf{WDJ182606.04+482911.30} has an orbital period of 39.82\,hr, with an orbitally-derived mass ratio $q=K_2/K_1=1.60\pm0.29$ and an atmospheric solution giving $q=0.87\pm0.12$. In this case, we found 3 periodogram peaks above the 4$\sigma$ false alarm probability threshold, but inspection of the RV curve for the 2 lower aliases shows a largely spurious result with only the RVs of the brighter star being well fit, and so they can be rejected. This leaves only the 39.82\,hr peak as a valid solution. The spectral fit for the atmospheric solution appears very good, though with a slightly over-predicted synthetic photometry in the blue. \textit{Gaia} photometry alone was used as there are two entries of the same source in PanSTARRS with reported magnitudes that slightly conflict, but when fitting just the spectra we obtain a very similar best-fit atmospheric solution. The reason for this large discrepancy in mass ratios eludes us, and we hope that future photometric/astrometric survey data releases or an improved precision on the orbital solution will aid in finding consistency in the future.

\textbf{WDJ183442.33$-$170028.00} had 3 solutions above the 4$\sigma$ false alarm probability. All fit the RVs from the hotter star well, but only the highest power peak also fits the RVs from the cooler WD, leaving one unique and valid solution. The source has an orbital period of 12.01\,hr, with an orbitally-derived mass ratio $q=K_2/K_1=0.807\pm0.468$ and the atmospheric solution represents $q=M_1/M_2=0.91\pm0.07$. The system is a unique case amongst the current DBL survey sample, having strong emission at H$\alpha$ which is not RV variable and likely originates from an unrelated hydrogen cloud given the location of the field. This severely jeopardises the precision of RVs from the secondary, leading to the high uncertainty on the orbital mass ratio. However, the found orbital solution is consistent with the atmospheric, and so we continue with the assumption that the atmospherically-deduced masses of both WDs are accurate. At approximately 0.4\,M$_\odot$ each, both stars are relatively low mass compared to the rest of the double-lined sample, and the merger time is $43\pm4$\,Gyr.

\textbf{WDJ212935.23+001332.26} has an orbital period of 15.01\,hr, with an orbitally-derived mass ratio $q=K_1/K_2=0.90\pm0.11$ and an atmospheric solution giving $q=1.00\pm0.10$. The Gaussians plus polynomial method was better and used here to extract RVs because of an inadequate atmospheric fit to the H$\alpha$ line cores.

\textbf{WDJ231404.30+552814.11} has an orbital period of 74.73\,hr, making it the largest orbital period of all system solutions presented in this work. The orbital mass ratio is $q=K_2/K_1=0.99\pm0.13$ while the atmospheric solution gives $q=1.74\pm0.12$ -- a striking discrepancy. With this, we decided to refit the atmospheric solution with a forced carbon-oxygen core relation for both stars, yet no combination of synthetic spectra agree with the orbital solution while performing hybrid fitting to the photometry, parallax and spectroscopy of the source. If we fit to the spectroscopy only and with a forced carbon-oxygen mass radius relationship, the fit appreciably improves in all Balmer lines besides H$\alpha$ and we are able to find an improved agreement with the orbital solution. The parameters we find are $T_1=13770\pm210$\,K, $T_2=8260\pm190$\,K, $\log(g_1)=7.86\pm0.05$\,dex, $\log(g_2)=7.64\pm0.04$\,dex, making $M_1=0.53\pm0.03$\,M$_\odot$ and $M_2=0.44\pm0.01$\,M$_\odot$. The distance (taken as the inverse of the parallax) is $105.0\pm0.3$\,pc for our fitted solution and $105.5\pm0.4$\,pc using the \textit{Gaia} DR3 parallax.  The mass ratio of the new atmospheric solution becomes $q=M_1/M_2=1.20\pm0.07$ and we find consistency between the atmospheric and orbital solutions. With that in mind, we encourage usage of this atmospheric solution which ignores the photometry and parallax of the source. In measuring RVs, we used the Gaussians plus polynomial model as it resulted in an improved accuracy and is independent of the best-fit synthetic spectra.

\subsection{Systems with period aliases and constraints}
\label{subsec:SystemsPeriodAliases}
Similar to before, here we address the systems where our RVs restrict the valid orbital solution of the binary by describing potential aliases. Presented are a further 5 of the 34 double-lined systems that we identified, plus details of a sixth source which was formerly considered a candidate double-lined and is now verified as double-lined. RVs for each source are presented in Appendix~\ref{appendix:RVsAliases}.

\textbf{WDJ000319.54+022623.28} has 17 RV measurements of each star recorded and has two peaks above the 4$\sigma$ false alarm probability, where it is difficult to separate which is correct. The peak of highest power is at a 14.55\,hr orbital period and the other at a 37.06\,hr orbital period. The relative flux difference between the H$\alpha$ line cores is large, meaning that, with approximately 3--5 more RV measurements closely separated in time, the true solution would be easily solvable in future work. Relatively uneven RV phase coverage leads to a large orbitally-derived mass ratio of about $q=K_2/K_1\approx1.9$ in both cases, but we promote more observations before drawing any further conclusions.

\textbf{WDJ080856.79+461300.08} has one unique peak above the 4$\sigma$ false alarm probability that poorly covers the orbital phase space, with an orbital period of 21.93\,hr. However, there is another peak with relatively high power at 19.48\,hr that appears suitable, leading us to categorise this as a system with aliases. Further smaller and similar height peaks exist in the periodogram in the range of 60--110\,hr, but they do not account for the short-term RV variability that is clearly noticed, with the peaks likely appearing because of the window function of our widely spaced observations. An orbital period of about 20\,hr seems the most likely solution, where the 21.93\,hr alias has $q=K_2/K_1=1.17\pm0.20$ and the 19.48\,hr has $q=1.15\pm0.33$, both consistent with that indicated from the atmospheric solution, which is $q=1.28\pm0.08$. The Gaussians plus polynomial method was used in obtaining RVs as the line core fit was significantly better.

\textbf{WDJ170120.99$-$191527.57} reveals the most compact orbital period of all systems presented in this work, but its true orbital period is not clear. Good orbital solutions are obtained above the 4$\sigma$ alarm probability for 4.95\,hr, 6.24\,hr and 4.10\,hr orbital periods, in order of periodogram peak power. A few out of phase points are witnessed in the RV curve for an 8.43\,hr period, which has about a third of the power as the other three peaks, but it is still significant. All of these solutions have an orbital mass ratio of $q\approx0.85$--$0.9$, while the atmospheric masses were found to be $0.673\pm0.024$\,M$_\odot$ and $0.544\pm0.022$\,M$_\odot$ (Section~\ref{subsec:atmosphericWDJ1701}), and with this the atmospheric $q=M_1/M_2=1.24\pm0.07$. 
For all of these orbital period aliases, the merger time of the system must be within a few gigayears.

\textbf{WDJ180115.37+721848.76} has 10 RV measurements in total, which are relatively few to solve an orbit, and with this the periodogram is inconclusive. The true alias of the DWD would be easy to resolve with just a couple more measurements even with a long time spacing given the clear shape of the H$\alpha$ line cores at a spectral resolution of $R=6310$. The atmospheric solution gives $q=0.89\pm0.05$, which aligns with the semi-amplitude ratio for the highest-power aliases, but for now we can only conclude that the orbital period is very likely between 6--40\,hr.

\textbf{WDJ221209.01+612906.96} has 9 RVs for each star. The spectral signature of each star is very similar, making it challenging to correctly assign to the hotter or cooler star at each epoch too. We tried all combinations of assignments possible, but without clear periodogram aliases appearing in all cases. Little can be said except that the orbital period is more than 3\,hr and the maximum possible period based on the binary mass function and the maximum perceived RV split between the stars is 13.5\,d. With the lack of certainty around the correct star at each epoch, we choose not to present the measured RVs for these systems.

\textbf{WDJ234929.57+102255.57} was originally categorised as a candidate double-lined DWD in \citet[][]{Munday2024DBL}. Further spectra clearly indicate a double-lined DWD, affirming the presumption. The periodogram of our 7 RV measurements give strong evidence in favour of an orbital period above 24\,hr, with prominent aliases of 27.49\,hr, 53.81\,hr, and 61.01\,hr. The latter two solutions have an orbital mass $q\approx1$, which would align well with the atmospheric solution, but the small number of spectra at hand are insufficient to draw a robust conclusion.

\begin{table*}
    \centering
    \caption{The measured orbital parameters for each system. For WDJ151109.90+404801.18, the period alias is resolved but some RV measurements are unclear, leading us not to present detailed orbital parameters or $T_0$.}
    \begin{tabular}{l|r|r|r|r|r|r}
    Name & Period & T$_0-2450000$ & K$_1$ & K$_2$ & $\gamma_1$ & $\gamma_2$\\
    & d & HJD, UTC & km\,s$^{-1}$ & km\,s$^{-1}$ & km\,s$^{-1}$ & km\,s$^{-1}$\\
    \hline
        WDJ002602.29$-$103751.86 & $0.9749777(58)$ & $8358.5743(60)$ & $110.1\pm5.8$ & $68.3\pm29.4$ & $8.4\pm3.4$ & $6.2\pm15.7$\\
        WDJ005413.14+415613.73 & $0.853294(70)$ & $8358.232(58)$ & $127.4\pm5.0$ & $99.0\pm8.9$ & $26.3\pm3.1$ & $51.0\pm6.1$  \\
         WDJ013812.93+444252.10 & $1.73324(13)$ & $8357.016(21)$ & $81.9\pm9.7$ & $97.5\pm 18.9$ & $25.1\pm3.6$ & $17.4\pm9.8$\\
         WDJ020847.22+251409.97 & $2.007653(28)$ & $8357.064(13)$ & $84.4\pm3.3$ & $71.9\pm 9.0$ & $49.1\pm 3.3$ & $36.2\pm 6.9$\\
         WDJ114446.16+364151.13 & $0.7492851(65)$ & $8645.273(12)$ & $108.6\pm 10.8$ & $111.7\pm 12.4$ & $6.5\pm 6.0$ & $4.5\pm 9.2$\\
         WDJ141625.94+311600.55 & $0.7867420(15)$ & $8589.519(25)$ & $125.5\pm 12.4$ & $86.9\pm 13.1$ & $26.0\pm 9.2$ & $38.3\pm 8.4$\\
         WDJ151109.90+404801.18 & $\approx0.97$--$0.99$ & - & - & - & - & - \\
         WDJ153615.83+501350.98  & $0.709254(48)$ & $8287.808(29)$ & $135.1\pm4.3$ & $89.2\pm9.9$ & $23.2\pm3.2$ & $44.5\pm7.0$\\
         WDJ160822.19+420543.44 & $0.8393438(36)$ & $8281.5690(18)$ & $92.6\pm2.5$ & $127.24\pm2.30$ & $-18.0\pm1.7$ & $-31.4\pm1.6$\\
         WDJ163441.85+173634.09  & $2.066817(44)$ & $8281.2062(80)$ & $79.6\pm1.7$ & $60.6\pm1.7$ & $16.2\pm1.8$ & $22.4\pm1.4$\\
         WDJ181058.67+311940.94 & $0.5931479(9)$ & $8587.6663(18)$ & $93.9\pm 2.0$ & $95.7\pm2.1$ & $50.0\pm1.5$ & $53.5\pm 1.6$\\
         WDJ182606.04+482911.30 & $1.659219(28)$ & $8358.620(25)$ & $72.7\pm9.3$ & $116.6\pm 14.9$ & $-17.3\pm 6.4$ & $-27.5\pm 10.2$\\
         WDJ183442.33$-$170028.00 & $0.5002200(39)$ & $8359.9874(54)$ & $194.2\pm29.1$ & $156.7\pm87.8$ & $25.1\pm25.2$ & $-18.6\pm73.2$\\
         WDJ212935.23+001332.26  & $0.6252288(96)$ & $8359.3171(76)$ & $108.2\pm 6.9$ & $97.5\pm 10.0$ & $31.4\pm 4.6$ & $32.6\pm 6.9$\\
         WDJ231404.30+552814.11 & 3.11386(15) & 8358.422(48) & $74.1\pm7.8$ & $73.2\pm6.1$ & $10.9\pm6.4$ & $-1.7\pm5.7$
         
    \end{tabular}
    \label{tab:my_label}
\end{table*}

\subsection{Follow-up of candidate double-lined DWDs and single-lined systems}
Further observations of a couple of single-lined targets were taken to probe RV variability when no other good targets were observable given the time and conditions of the night, but all of them resulted in no firm detection of binarity. These consist of three spectra of WDJ185640.86+120844.61 and ten spectra of WDJ192817.81+354442.60. The extra RV measurements for these systems, as well as all RVs for the single-lined systems presented in \citet{Munday2024DBL}, are presented in Appendix~\ref{appendix:RVsSB1}.

\section{Discussion}
With many new orbital solutions presented, we turn our attention towards making comparisons with the full sample of published DWD binaries\footnote{https://github.com/JamesMunday98/CloseDWDbinaries \citep[][]{Munday2024DBL}}. We emphasise again that the choice of which double-lined systems were followed up for time-series RVs was purely based upon which were observable for the allocated telescope nights and for systems where both stars would be separable at the slightly lower spectral resolution of the INT observations. Hence, the sampling of systems from the full list of double-lined DWDs in the DBL survey is relatively random, including the outlier in total mass WDJ181058.67+311940.94 which was the highest total mass system discovered.

\begin{figure}
    \centering
    \includegraphics[width=\columnwidth,trim={0.85cm 0.35cm 2.02cm 1.75cm}, clip]{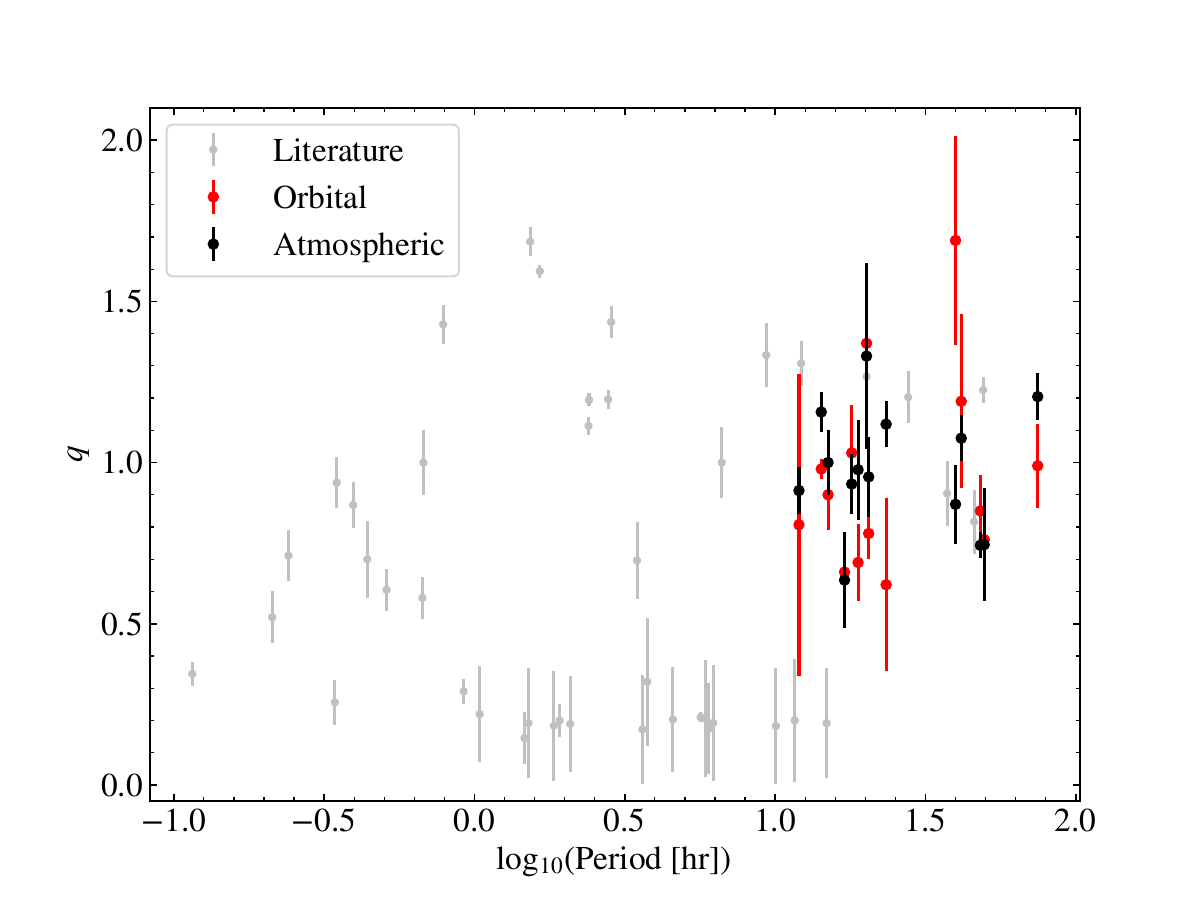}
    \caption{A comparison of the orbital and atmospheric mass ratios against the orbital period. The mass ratios of other previously studied DWDs are included in grey to show the overall observed mass ratio versus period distribution of DWDs. Mass ratios of the double-lined DWDs found in the DBL survey are limited to approximately $q=0.5$--$2.0$.}
    \label{fig:PeriodVsQ}
\end{figure}

\begin{figure}
    \centering
    \includegraphics[width=\columnwidth,trim={1.07cm 0.4cm 1.95cm 1.75cm}, clip]{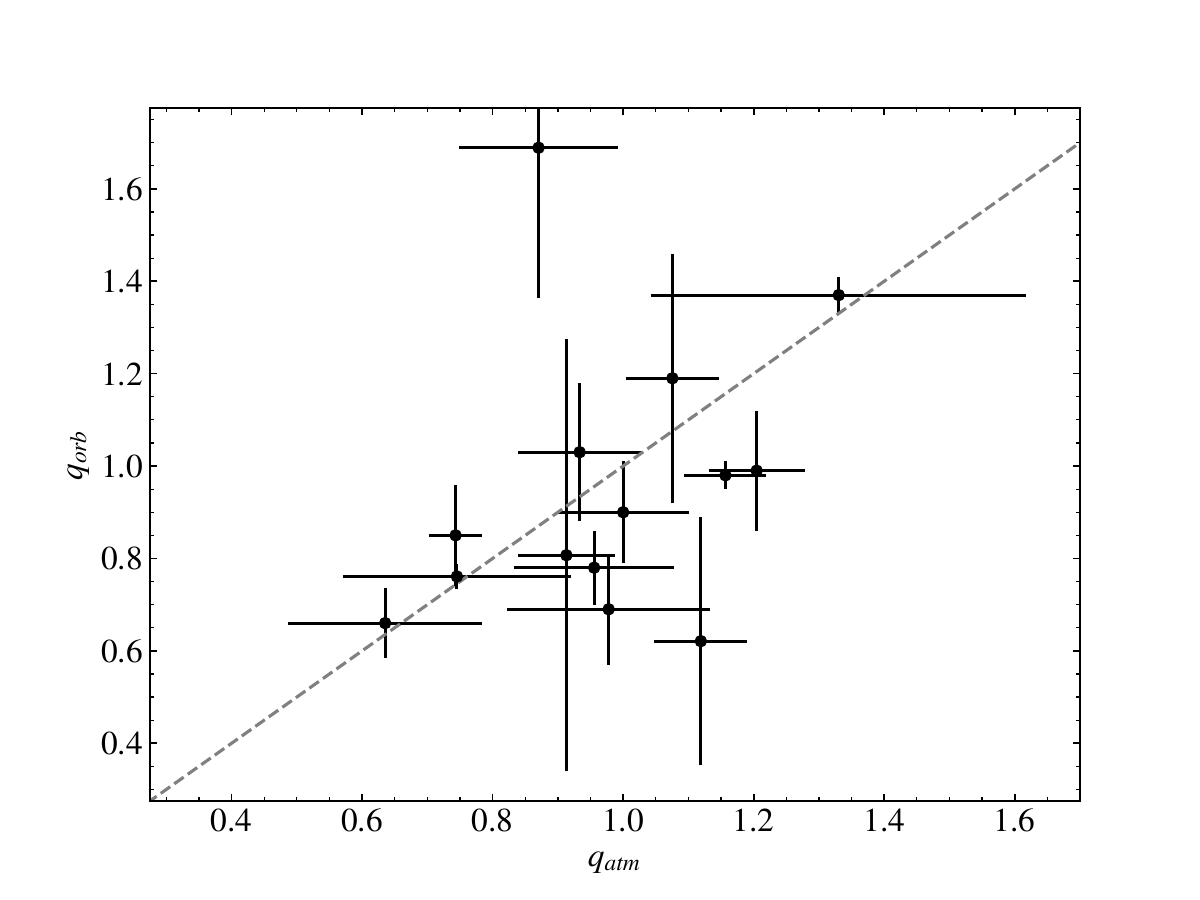}
    \caption{A comparison of the mass ratios deduced from the orbital solution and the fitting of synthetic spectra to obtain atmospherically determined masses. The dashed line represents $q_\text{orb}=q_\text{atm}$.}
    \label{fig:qorb_vs_qatm}
\end{figure}

Fig.~\ref{fig:PeriodVsQ} shows the orbital period and mass ratio distribution of the observed sample. As is clear from the previous section and Fig.~\ref{fig:PeriodVsQ}, these DWD binaries group around an orbital mass ratio of $q=1$. This is an unsurprising observational bias of the survey, given that two similar stars are easiest to detect as a double-lined source DWD. The newly discovered systems populate a period space that before was sparsely sampled. Their range of periods aligns well with predictions from recent synthetic populations, which forecast an abundance of $q\approx1$ DWDs at certain periods \citep[][]{Ge2010, Toonen2012type1aCommonEnvelope, Ge2015, Ge2020, Li2023}. In the vast majority of cases, we see that the orbital and atmospheric mass ratios agree well (Figs.~\ref{fig:PeriodVsQ} and \ref{fig:qorb_vs_qatm}), emphasising the general accuracy of the masses of double-lined DWDs obtained from identification spectra alone when fitting full-visible spectroscopy in combination with absolute photometry and precise parallaxes, although the discrepancy varies on a case by case basis. This discrepancy is especially relevant for systems where at least one star has a mass close to 0.45\,M$_\odot$, which is the mass we used to transition between a carbon-oxygen or a helium core mass radius relationship in the atmospheric parameter fitting.

\begin{figure}
    \centering
    \includegraphics[width=\columnwidth,clip,trim={1.025cm 0.4cm 2cm 1.75cm}]{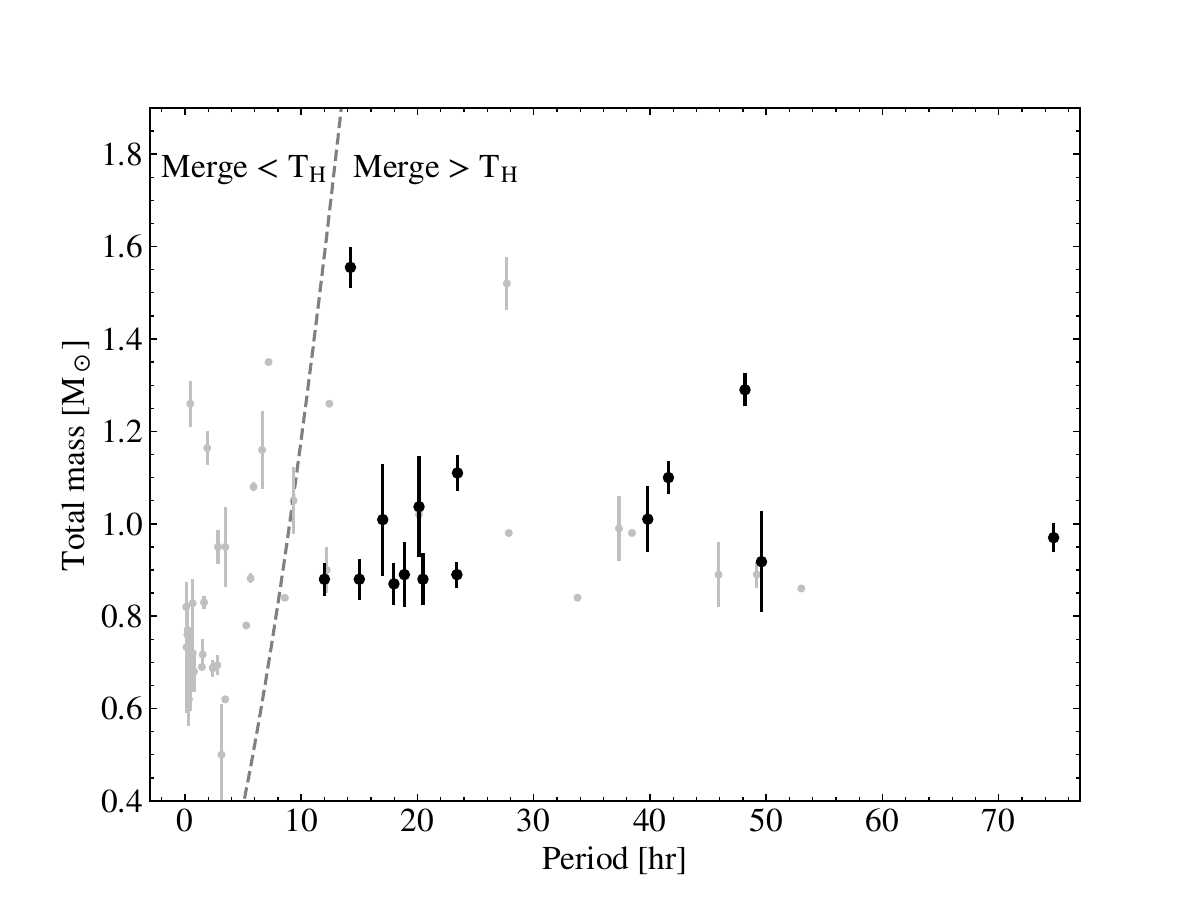}
\caption{The total mass versus orbital period distribution of the observed sample of DWDs. The points in gray circles are systems in the literature that are eclipsing or double-lined, whereas the black crosses are systems systems with orbital solutions reported in this work. The dashed grey line is, for two stars of equal mass, the boundary for a binary merging within a Hubble time (T$_\text{H}$). One double-lined DWD \citep[PG1115+166,][]{2002ApJ...566.1091B} is not plotted for clarity because of its approximately 30\,d orbital period \citep[][]{2002MNRAS.334..833M}. The total masses of the targets that fall in the DBL survey are restricted to approximately 0.8--2.0\,M$_\odot$ because of the Hertzsprung-Russell diagram cuts imposed.}
    \label{fig:PeriodVsTotalMass}
\end{figure}

\begin{figure}
    \centering
    \includegraphics[width=\columnwidth, clip, trim={1cm 0.4cm 2cm 1.75cm}]{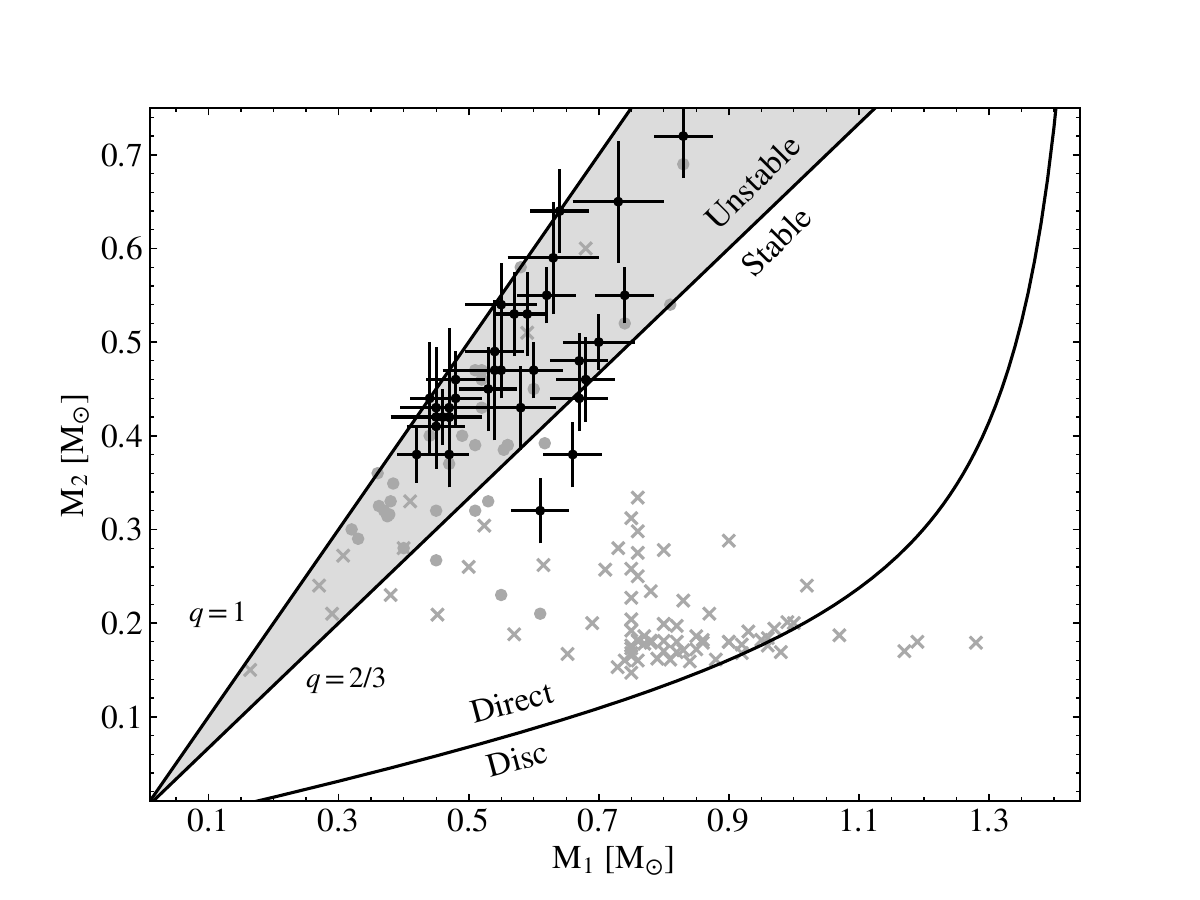}
    \caption{The location of all DBL survey double-lined systems (black) versus other systems in the literature (grey) when comparing a future stable versus unstable mass transfer scenario and whether disc or direct accretion will eventually take place. The literature points in circles are double-lined DWDs and those in crosses are single-lined DWDs. Direct impact accretion occurs when the trajectory of the accretion stream collides with the surface of the accretor, serving as an angular momentum sink that speeds up the inspiral of a binary, and observationally recognised by strong electromagnetic pulses on the orbital period.}
    \label{fig:StableUnstable}
\end{figure}

With our selection of DWDs having similar masses for both stars, the total masses of the DBL systems characterised in this work mostly fall around 1.0\,M$_\odot$, shown in Fig.~\ref{fig:PeriodVsTotalMass}. In comparison to other double-lined or eclipsing systems in the literature (which, generally speaking, are those with accurate masses for both components), the systems investigated in this paper double the number of compact DWDs with an orbital period above 10\,hr. This much better reflects the expectations of synthetic populations. All of these newly-characterised systems will merge in more than a Hubble time, with the one notable exception being WDJ170120.99$-$191527.57, having an unclear orbital solution. With a mass ratio close to one, unstable mass transfer will likely cause a merger in the future of these binaries, leading to a R~CrB star or larger mass, single WD. We show how close each system is to the boundary of stable and unstable mass transfer in Fig.~\ref{fig:StableUnstable}. WDJ181058.67+311940.94 will also initiate unstable mass transfer, but its total mass significantly exceeds the Chandrasekhar mass limit, inevitably giving rise to a type Ia supernova explosion before the event of merger \citep[][]{Munday2025nature}.

\begin{figure}
    \centering
    \includegraphics[width=\columnwidth,clip,trim={1cm 0.4cm 2cm 1.8cm}]{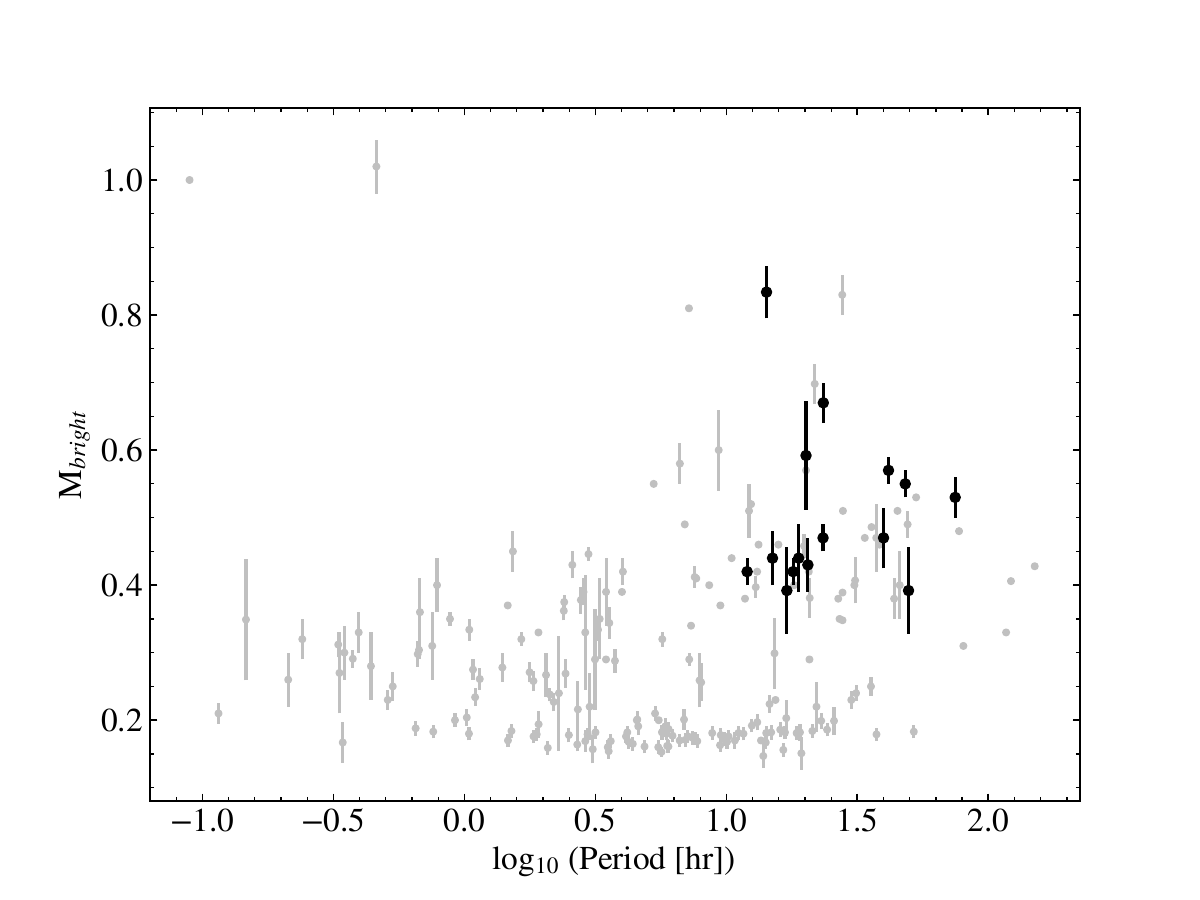}
    \caption{The mass of the brighter star in DWD binaries compared with the orbital period. The clumping of systems with M$_\textrm{bright}=0.2$\,M$_\odot$ are ELM binaries that are confirmed to be DWDs stemming largely from the ELM survey \citep[][]{Brown2020elmNorthFinal, Kosakowski2023elmSouth} which primarily form from the mass transfer sequence of a common envelope followed by stable Roche lobe overflow. The brightest absolute magnitude possible for a target found in the DBL survey is similar to a 0.4\,M$_\odot$ DA WD evolutionary sequence for a helium core mass-radius relationship, making 0.35--0.4\,M$_\odot$ the minimum mass of the brighter component.}
    \label{fig:MbrightPorb}
\end{figure}

\begin{figure}
    \centering
    \includegraphics[width=\columnwidth,clip,trim={1cm 0.4cm 1.75cm 1.75cm}]{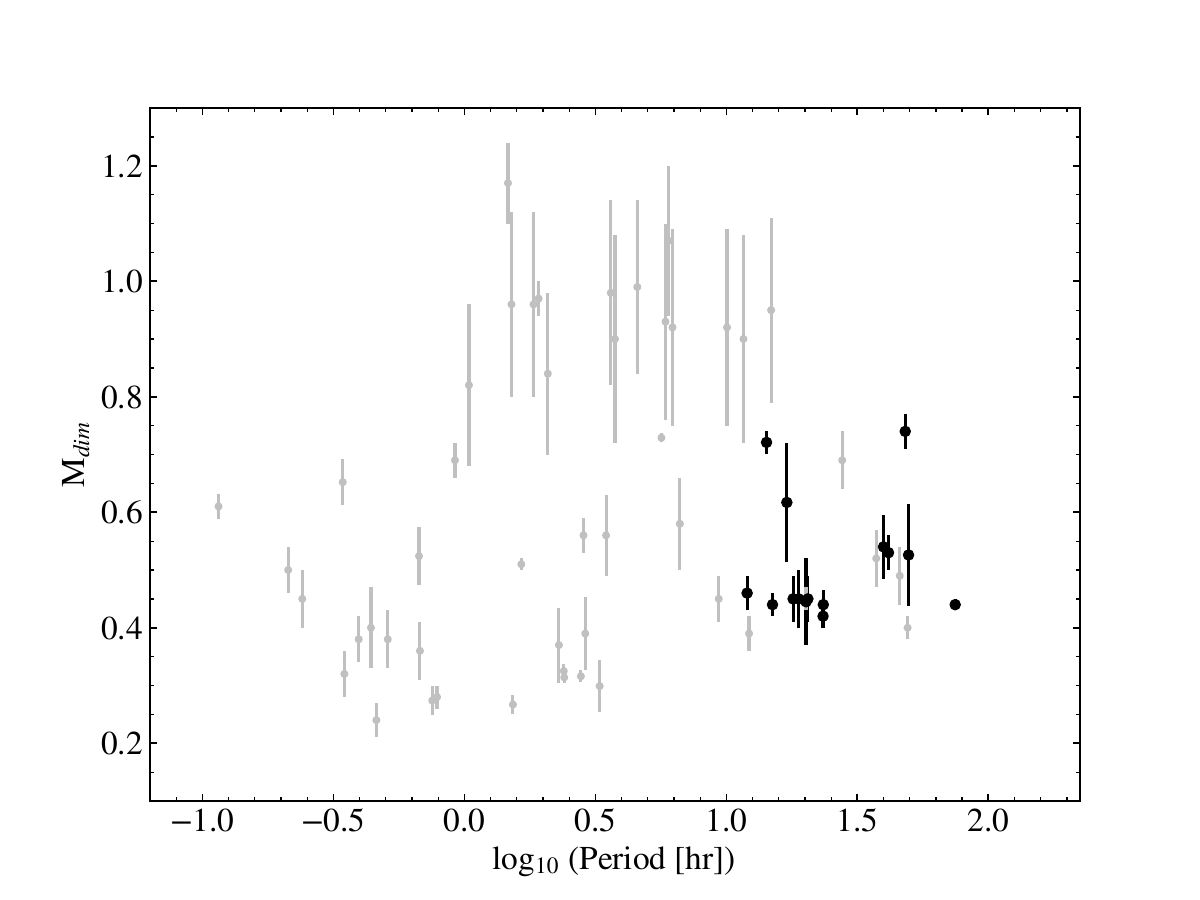}
    \caption{The same as Fig.~\ref{fig:MbrightPorb}, but for the dimmer star of the binary (M$_\text{dim}$). We limit plotting of the literature systems where the fractional error of the mass of the dimmer star is below 20\% for the purpose of clarity, which appreciably removes cases around 0.8--1.0\,M$_\odot$.}
    \label{fig:MdimPorb}
\end{figure}

Fig.~\ref{fig:MbrightPorb} again shows the orbital periods of the DWD sample but compared with the mass of the brighter star, while Fig.~\ref{fig:MdimPorb} depicts the mass of the dimmer star. A trend of an increasing mass of the brighter star with an increasing orbital period is slightly evident for the presented DBL survey sources (besides the one case of WDJ181058.67+311940.94). This likely appears because mass transfer initiated later for the systems showing longer orbital periods after the common envelope phase, allowing the WD progenitor to grow to higher core masses. The trend here can be a powerful tool in constraining an empirical relationship for the final separation of DWDs exiting the common envelope phase, and with this a correlation with the core mass of the former giant star, but it is difficult to draw strong conclusions with the sample size.

\begin{figure}
    \centering
    \includegraphics[width=\columnwidth,clip,trim={1cm 0.35cm 1.65cm 1.6cm}]{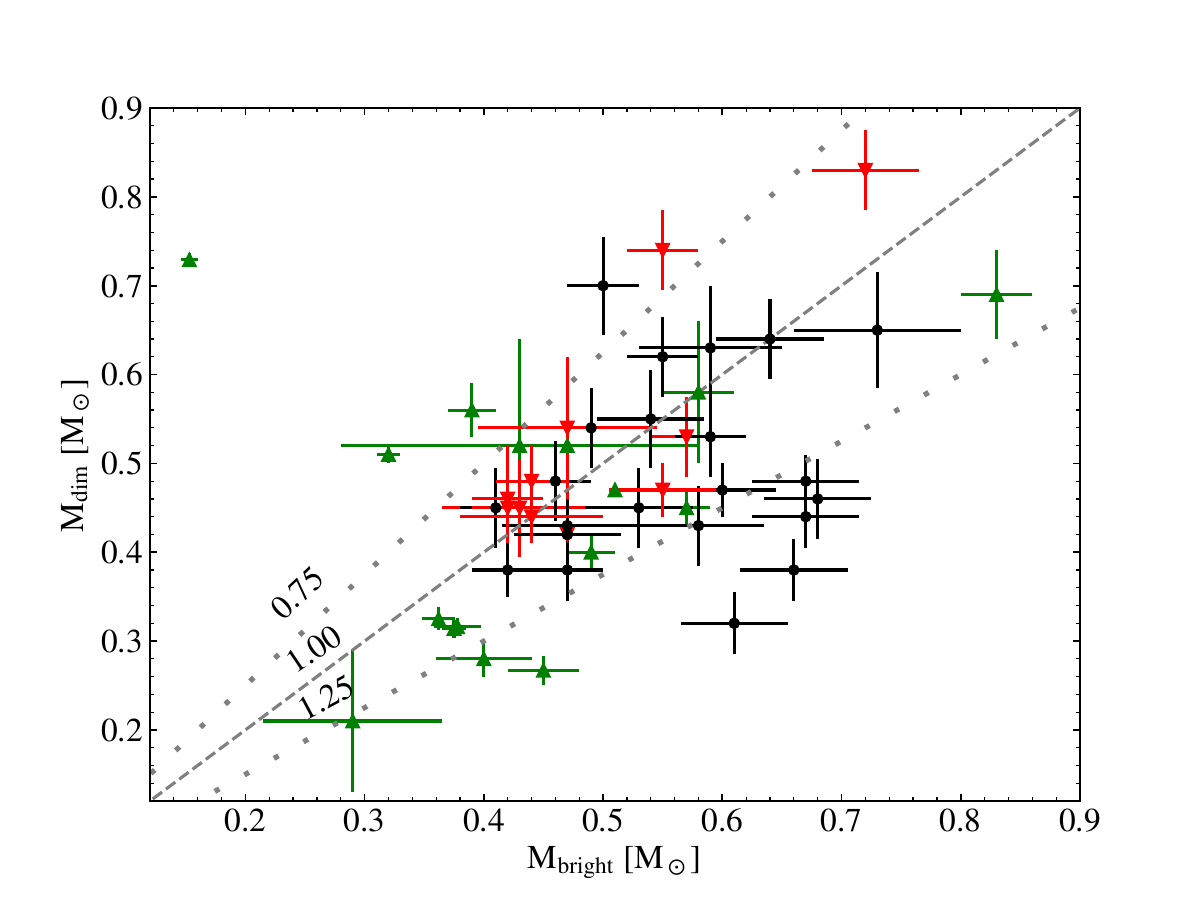}
    \caption{A comparison of the mass of the brighter (more luminous) and dimmer component for well-characterised DWDs. The downward facing and red triangles are the systems with complete orbital solutions discussed in Section~\ref{subsec:resultsSolvedPeriods}. Black points are all other systems presented in the DBL~\Romannum{1} study. The upward facing green triangles are other eclipsing or double-lined systems found in the literature. In dotted lines are ratios of $\text{M}_\text{bright}/\text{M}_\text{dim}=0.75, 1.00$ and 1.25.}
    \label{fig:MbrightMdim}
\end{figure}

Additionally, a graphical representation of the masses of the brighter and dimmer stars in the orbitally solved binaries from this paper can be found in Fig.~\ref{fig:MbrightMdim}. A concentration of systems can be seen following the trend $\text{M}_\text{bright}/\text{M}_\text{dim}=1.25$. Overall, about one-quarter of the 34 double-lined systems give a brighter star that is more massive. This situation can arise from the ``formation reversal channel'' \citep[][]{Toonen2012type1aCommonEnvelope}, in which the first mass transfer phase initially forms a helium star that then cools to become a WD, thus delaying the star's arrival to the cooling sequence and making it appear younger than the later-formed companion WD. The large fraction of these systems in our survey could be an artifact of the magnitude limit 
employed in the DBL survey, rather than being reflective of the full DWD population, since the first-formed WD remains brighter for longer in this evolutionary channel.

\section{Conclusions}
We have presented a continuation of the DBL survey supplying orbital solutions for 15 double-lined DWD binaries and have narrowed down the viable period aliases for a further 6 DWDs. These systems cover a wide range of orbital periods for compact DWD binaries, spreading between $P=5$--75\,hr. We directly analysed the compatibility of the mass ratio derived from the orbital solutions with the mass ratio determined from spectral fitting, finding a good agreement with the values deduced from identification spectra alone. This hence emphasises the potential power of hybrid (combined photometric and spectroscopic) fitting in deducing a mass distribution of double-lined DWDs in wide-scale searches through medium-resolution spectra.

Ultimately, we have taken the first steps in obtaining a mass-period distribution of double-lined DWDs that lie within the selection criteria of the DBL survey, outlined in \citet[][]{Munday2024DBL}. This will be a powerful tool to calibrate synthetic population models of DWDs when exploring systems where the mass of the brighter WD exceeds approximately 0.4\,M$_\odot$. Future work will aspire to build a larger and more robust sample towards this goal, both in terms of orbital characterisation and of identification of new double-lined DWDs.

\section*{Acknowledgements}
We thank Yuri Beletsky for their support in obtaining spectra from MagE on the Magellan Baade telescope. JM was supported by funding from a Science and Technology Facilities Council (STFC) studentship. IP acknowledges support from a Royal Society University Research Fellowship (URF\textbackslash R1\textbackslash 231496). This research received funding from the European Research Council under the European Union’s Horizon 2020 research and innovation programme number 101002408 (MOS100PC). DJ acknowledges support from the Agencia Estatal de Investigaci\'on del Ministerio de Ciencia, Innovaci\'on y Universidades (MCIU/AEI) under grant ``Nebulosas planetarias como clave para comprender la evoluci\'on de estrellas binarias'' and the European Regional Development Fund (ERDF) with reference PID-2022-136653NA-I00 (DOI:10.13039/501100011033). DJ also acknowledges support from the Agencia Estatal de Investigaci\'on del Ministerio de Ciencia, Innovaci\'on y Universidades (MCIU/AEI) under grant ``Revolucionando el conocimiento de la evoluci\'on de estrellas poco masivas'' and the the European Union NextGenerationEU/PRTR with reference CNS2023-143910 (DOI:10.13039/501100011033). MK acknowledges support from the NSF under grant AST-2205736 and the NASA under grants 80NSSC22K0479, 80NSSC24K0380, and 80NSSC24K0436. By NASA through the NASA Hubble Fellowship grant HST-HF2-51527.001-A awarded by the Space Telescope Science Institute, which is operated by the Association of Universities for Research in Astronomy, Inc., for NASA, under contract NAS5-26555. ST acknowledges support from the Netherlands Research Council NWO (VIDI 203.061 grant). For the purpose of open access, the authors have applied a creative commons attribution (CC BY) licence to any author accepted manuscript version arising.

The Isaac Newton Telescope and the William Herschel Telescope are operated on the island of La Palma by the Isaac Newton Group of Telescopes in the Spanish Observatorio del Roque de los Muchachos of the Instituto de Astrofísica de Canarias. Based on observations made with the Nordic Optical Telescope, owned in collaboration by the University of Turku and Aarhus University, and operated jointly by Aarhus University, the University of Turku and the University of Oslo, representing Denmark, Finland and Norway, the University of Iceland and Stockholm University at the Observatorio del Roque de los Muchachos, La Palma, Spain, of the Instituto de Astrof\'isica de Canarias. The data presented here were obtained in part with ALFOSC, which is provided by the Instituto de Astrof\'isica de Andaluc\'ia (IAA) under a joint agreement with the University of Copenhagen and NOT. This paper includes data gathered with the 6.5 meter Magellan Telescopes located at Las Campanas Observatory, Chile.

Based on observations obtained at the international Gemini Observatory, a program of NSF's NOIRLab, which is managed by the Association of Universities for Research in Astronomy (AURA) under a cooperative agreement with the National Science Foundation on behalf of the Gemini Observatory partnership: the National Science Foundation (United States), National Research Council (Canada), Agencia Nacional de Investigaci\'{o}n y Desarrollo (Chile), Ministerio de Ciencia, Tecnolog\'{i}a e Innovaci\'{o}n (Argentina), Minist\'{e}rio da Ci\^{e}ncia, Tecnologia, Inova\c{c}\~{o}es e Comunica\c{c}\~{o}es (Brazil), and Korea Astronomy and Space Science Institute (Republic of Korea).
\section*{Data Availability}
Measured RVs are given in the Appendix~\ref{appendix:RVs}. Raw spectra are obtainable through the respective telescope's public data archive. Reduced spectra will be made available upon request to the lead author.



\bibliographystyle{mnras}
\bibliography{mnras_template} 




\appendix
\section{Spectroscopic fit to WDJ170120.99$-$191527.57}
\begin{figure}
    \centering
    \includegraphics[width=\columnwidth, clip, trim={0cm 0cm 1.5cm 1.75cm}]{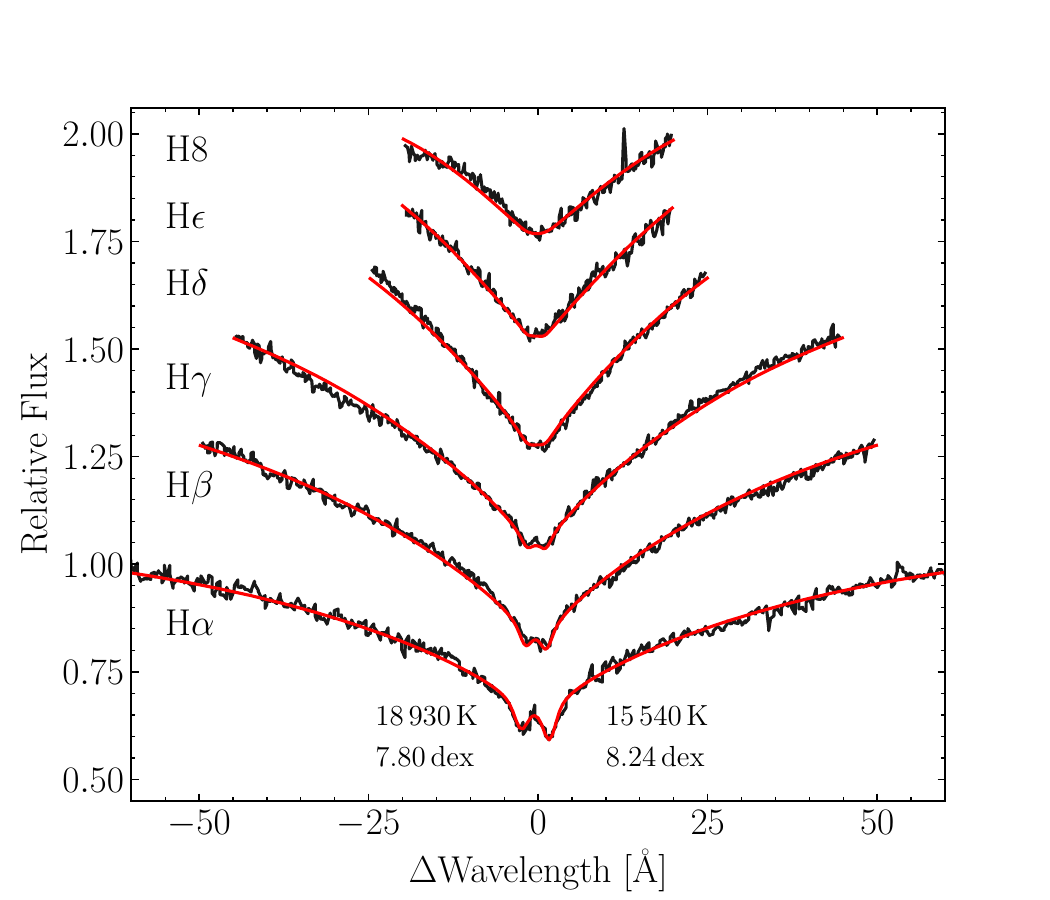}
    \caption{The spectroscopic fit (red) overlaid on a Magellan/MagE spectrum (black) of WDJ170120.99$-$191527.57 from the Balmer lines H$\alpha$ to H8. The atmospheric parameters for the two stars are labelled, where here the blue-shifted star is hotter and less massive.}
    \label{fig:wdj1701_fit}
\end{figure}

With the new and higher quality data described in Section~\ref{subsec:atmosphericWDJ1701}, we present in Fig.~\ref{fig:wdj1701_fit} the atmospheric fit to this double-lined DWD.

\section{Radial velocities of fully solved double lined systems}
\label{appendix:RVs}
Found in Tables~\ref{tab:AppendixRVsWDJ002602}--\ref{tab:AppendixRVsWDJ231404} are the observed RVs of every source with a solved orbit or with a confined list of period alias.

\begin{table}
\centering
\caption{RVs for the target WDJ002602.29$-$103751.86}
\begin{tabular}{r|r|r|r|r} \\
HJD$-$2450000 & $\Delta$RV$_1$ & RV$_1$ & RV$_2$ & $\Delta$RV$_2$\\
\text{[d]} & [km\,s$^{-1}$] & [km\,s$^{-1}$] & [km\,s$^{-1}$] & [km\,s$^{-1}$] \\

\hline
8363.05908 & +85.8 & $\pm$4.2 & $-$38.3 & $\pm$14.1\\
8366.06724 & +112.8 & $\pm$4.2 & $-$55.0 & $\pm$23.2\\
8366.08471 & +112.3 & $\pm$4.4 & $-$69.9 & $\pm$52.1\\
8726.12643 & $-$4.5 & $\pm$11.0 & $-$5.2 & $\pm$40.5\\
8726.14740 & +6.3 & $\pm$12.9 & +75.5 & $\pm$53.8\\
8726.17061 & $-$24.0 & $\pm$8.8 & +39.7 & $\pm$42.2\\
8726.19158 & $-$41.6 & $\pm$12.0 & +87.0 & $\pm$43.9\\
8726.21485 & $-$48.9 & $\pm$10.8 & +70.9 & $\pm$30.6\\
8726.23582 & $-$65.3 & $\pm$9.7 & +60.6 & $\pm$50.5\\
8729.08342 & $-$5.3 & $\pm$9.5 & $-$21.1 & $\pm$48.0\\
8729.10441 & $-$28.0 & $\pm$11.7 & +41.8 & $\pm$33.7\\
8730.15279 & $-$65.6 & $\pm$9.2 & +122.8 & $\pm$54.7\\
8730.17377 & $-$83.9 & $\pm$10.0 & +19.9 & $\pm$51.9\\
8730.19727 & $-$89.3 & $\pm$9.5 & +18.3 & $\pm$45.0\\
8732.04642 & $-$25.7 & $\pm$16.2 & +18.8 & $\pm$40.4\\
8750.02581 & +14.4 & $\pm$10.9 & +82.6 & $\pm$19.4\\
8750.04678 & +36.5 & $\pm$10.3 & $-$67.6 & $\pm$36.3\\
10699.82972 & $-$79.5 & $\pm$7.0 & +54.8 & $\pm$25.6\\
10699.85443 & $-$59.3 & $\pm$7.9 & +20.7 & $\pm$29.2\\
10700.83261 & $-$64.9 & $\pm$10.3 & +9.1 & $\pm$42.5\\
10700.85732 & $-$61.4 & $\pm$17.1 & +40.2 & $\pm$43.2\\
10701.83575 & $-$53.3 & $\pm$8.7 & +48.4 & $\pm$45.0\\
\end{tabular}
\label{tab:AppendixRVsWDJ002602}
\end{table}

\begin{table}
\centering
\caption{RVs for the target WDJ005413.14+415613.73}
\begin{tabular}{r|r|r|r|r} \\
HJD$-$2450000 & $\Delta$RV$_1$ & RV$_1$ & RV$_2$ & $\Delta$RV$_2$\\
\text{[d]} & [km\,s$^{-1}$] & [km\,s$^{-1}$] & [km\,s$^{-1}$] & [km\,s$^{-1}$] \\

\hline
8359.22546 & $-$29.3 & $\pm$2.9 & +88.0 & $\pm$4.9\\
8361.16421 & +148.3 & $\pm$3.3 & $-$43.3 & $\pm$3.1\\
8723.09082 & +124.1 & $\pm$11.9 & $-$18.2 & $\pm$16.6\\
8723.11179 & +120.5 & $\pm$5.1 & $-$23.4 & $\pm$10.5\\
8723.13841 & +87.9 & $\pm$6.2 & +2.9 & $\pm$8.0\\
8723.15938 & +64.9 & $\pm$11.4 & +8.6 & $\pm$13.5\\
8723.18289 & +56.6 & $\pm$7.5 & +21.0 & $\pm$12.3\\
8723.20386 & +37.5 & $\pm$5.6 & +38.0 & $\pm$12.9\\
8724.23697 & $-$96.5 & $\pm$4.0 & +159.2 & $\pm$6.4\\
8724.25795 & $-$95.3 & $\pm$33.9 & +135.6 & $\pm$35.6\\
8725.01340 & $-$51.5 & $\pm$4.5 & +127.7 & $\pm$9.1\\
8725.03438 & $-$65.8 & $\pm$3.9 & +125.8 & $\pm$7.3\\
8725.05750 & $-$72.3 & $\pm$4.1 & +129.9 & $\pm$8.2\\
8725.07848 & $-$90.0 & $\pm$4.9 & +135.8 & $\pm$7.9\\
8750.95723 & +56.5 & $\pm$15.6 & $-$34.0 & $\pm$24.9\\
8751.01084 & +54.4 & $\pm$25.1 & +3.9 & $\pm$16.3\\
8751.08709 & +126.5 & $\pm$23.7 & $-$28.0 & $\pm$18.1\\
8751.10808 & +161.5 & $\pm$21.9 & +4.1 & $\pm$26.1\\
8751.12909 & +136.3 & $\pm$14.1 & $-$17.4 & $\pm$21.9\\
\end{tabular}
\label{tab:AppendixRVsWDJ005413}
\end{table}

\begin{table}
\centering
\caption{RVs for the target WDJ013812.93+444252.10}
\begin{tabular}{r|r|r|r|r} \\
HJD$-$2450000 & $\Delta$RV$_1$ & RV$_1$ & RV$_2$ & $\Delta$RV$_2$\\
\text{[d]} & [km\,s$^{-1}$] & [km\,s$^{-1}$] & [km\,s$^{-1}$] & [km\,s$^{-1}$] \\

\hline
8359.20729 & $-$1.6 & $\pm$2.2 & +39.5 & $\pm$7.3\\
8361.19638 & $-$58.9 & $\pm$3.0 & +110.3 & $\pm$6.0\\
8361.21038 & $-$53.9 & $\pm$3.3 & +115.9 & $\pm$7.5\\
8723.23426 & $-$15.9 & $\pm$12.0 & +84.4 & $\pm$10.6\\
8723.25524 & $-$74.6 & $\pm$31.3 & +57.8 & $\pm$32.3\\
8724.04965 & +50.6 & $\pm$7.8 & +29.4 & $\pm$17.2\\
8724.07063 & +67.5 & $\pm$9.0 & $-$12.4 & $\pm$11.8\\
8724.09390 & +70.7 & $\pm$7.4 & $-$4.1 & $\pm$26.9\\
8724.11487 & +62.5 & $\pm$8.2 & $-$41.6 & $\pm$18.2\\
8724.14170 & +81.9 & $\pm$6.3 & $-$51.2 & $\pm$24.3\\
8724.16268 & +60.9 & $\pm$9.6 & $-$60.2 & $\pm$18.2\\
8724.18599 & +82.2 & $\pm$11.2 & $-$38.7 & $\pm$26.4\\
8724.20696 & +94.1 & $\pm$8.9 & $-$55.6 & $\pm$17.5\\
8726.98848 & $-$55.5 & $\pm$8.7 & +75.4 & $\pm$33.2\\
8727.00947 & $-$57.3 & $\pm$11.4 & +96.5 & $\pm$14.2\\
8731.15622 & +90.3 & $\pm$5.5 & $-$76.4 & $\pm$13.4\\
\end{tabular}
\label{tab:AppendixRVsWDJ013812}
\end{table}

\begin{table}
\centering
\caption{RVs for the target WDJ020847.22+251409.97}
\begin{tabular}{r|r|r|r|r} \\
HJD$-$2450000 & $\Delta$RV$_1$ & RV$_1$ & RV$_2$ & $\Delta$RV$_2$\\
\text{[d]} & [km\,s$^{-1}$] & [km\,s$^{-1}$] & [km\,s$^{-1}$] & [km\,s$^{-1}$] \\

\hline
8359.17717 & +128.8 & $\pm$1.6 & $-$30.3 & $\pm$7.1\\
8361.18090 & +130.1 & $\pm$1.7 & $-$31.9 & $\pm$8.1\\
8362.23391 & $-$27.0 & $\pm$1.7 & +105.8 & $\pm$6.2\\
8364.25678 & $-$24.3 & $\pm$1.7 & +109.5 & $\pm$5.3\\
8732.06703 & +65.9 & $\pm$6.1 & +3.2 & $\pm$28.7\\
8732.07759 & +70.9 & $\pm$3.8 & +1.0 & $\pm$23.8\\
8732.12265 & +75.4 & $\pm$5.2 & +13.2 & $\pm$16.4\\
8732.13321 & +87.0 & $\pm$5.0 & $-$13.4 & $\pm$20.0\\
8732.14619 & +84.8 & $\pm$5.5 & $-$16.3 & $\pm$21.0\\
8732.15675 & +89.5 & $\pm$4.5 & $-$2.3 & $\pm$11.1\\
8733.13824 & +45.9 & $\pm$16.3 & $-$37.0 & $\pm$99.9\\
8733.14880 & +10.8 & $\pm$13.4 & +92.9 & $\pm$85.1\\
8733.16157 & +6.9 & $\pm$11.4 & +71.1 & $\pm$77.8\\
8733.17214 & +19.1 & $\pm$12.1 & +73.0 & $\pm$66.7\\
8734.11706 & +70.2 & $\pm$3.1 & +15.2 & $\pm$11.8\\
8734.13803 & +82.5 & $\pm$2.6 & $-$7.4 & $\pm$9.8\\
8734.15596 & +90.2 & $\pm$3.0 & $-$0.1 & $\pm$12.0\\
8734.16652 & +92.4 & $\pm$4.1 & $-$1.8 & $\pm$11.4\\
8734.17855 & +95.4 & $\pm$5.0 & +7.4 & $\pm$13.2\\
8750.15523 & +77.4 & $\pm$2.8 & $-$17.6 & $\pm$7.7\\
8750.16578 & +76.2 & $\pm$3.1 & $-$23.8 & $\pm$17.6\\
8750.17634 & +73.2 & $\pm$3.9 & $-$31.9 & $\pm$12.5\\
8750.18690 & +78.7 & $\pm$3.5 & $-$6.7 & $\pm$13.8\\
8751.03389 & +43.0 & $\pm$8.9 & +93.8 & $\pm$47.8\\
8751.04445 & +64.2 & $\pm$8.8 & +30.5 & $\pm$55.1\\
8751.05501 & +72.3 & $\pm$9.0 & $-$27.2 & $\pm$27.0\\
8751.06557 & +44.2 & $\pm$9.0 & +85.1 & $\pm$44.2\\
10214.06255 & +125.6 & $\pm$3.7 & $-$45.3 & $\pm$15.1\\
10214.06848 & +142.2 & $\pm$5.0 & $-$54.9 & $\pm$22.9\\
10214.07441 & +130.6 & $\pm$4.5 & $-$15.2 & $\pm$13.4\\
10214.08034 & +135.1 & $\pm$4.2 & $-$1.9 & $\pm$16.1\\
10214.18215 & +131.1 & $\pm$4.9 & $-$34.3 & $\pm$15.1\\
10214.18924 & +134.8 & $\pm$5.0 & $-$44.2 & $\pm$12.7\\
10214.19690 & +126.2 & $\pm$3.7 & $-$37.0 & $\pm$36.7\\
10214.20399 & +136.7 & $\pm$4.1 & $-$13.8 & $\pm$21.2\\
10233.95335 & +88.6 & $\pm$4.0 & $-$35.1 & $\pm$14.1\\
10233.96102 & +110.9 & $\pm$4.2 & +1.2 & $\pm$17.2\\
10233.96813 & +114.4 & $\pm$3.9 & +11.8 & $\pm$11.1\\
10233.97523 & +99.2 & $\pm$4.0 & $-$48.0 & $\pm$13.2\\
10233.98290 & +104.2 & $\pm$3.6 & +0.3 & $\pm$11.7\\
10234.01358 & +128.1 & $\pm$3.9 & +0.4 & $\pm$18.3\\
10234.02069 & +114.2 & $\pm$3.8 & $-$19.2 & $\pm$8.8\\
10234.02780 & +121.4 & $\pm$4.5 & $-$7.2 & $\pm$11.6\\
10234.03535 & +130.3 & $\pm$4.8 & $-$41.4 & $\pm$41.7\\
10234.08485 & +130.7 & $\pm$4.8 & $-$41.7 & $\pm$15.7\\
10234.09195 & +120.0 & $\pm$3.8 & $-$64.7 & $\pm$12.6\\
10234.09906 & +153.5 & $\pm$4.6 & $-$7.3 & $\pm$19.8\\
10234.10662 & +131.2 & $\pm$3.8 & $-$45.5 & $\pm$17.5\\
10234.11390 & +131.0 & $\pm$4.4 & $-$18.7 & $\pm$14.2\\
10234.16524 & +131.2 & $\pm$4.3 & $-$64.6 & $\pm$18.0\\
10234.17235 & +137.9 & $\pm$4.4 & $-$22.4 & $\pm$14.5\\
10234.17946 & +125.4 & $\pm$4.1 & $-$38.1 & $\pm$12.6\\
10234.18714 & +140.1 & $\pm$5.2 & +7.7 & $\pm$15.9\\
10234.19424 & +116.5 & $\pm$3.8 & $-$25.4 & $\pm$10.3\\
\end{tabular}
\label{tab:AppendixRVsWDJ020847}
\end{table}

\begin{table}
\centering
\caption{RVs for the target WDJ114446.16+364151.13}
\begin{tabular}{r|r|r|r|r} \\
HJD$-$2450000 & $\Delta$RV$_1$ & RV$_1$ & RV$_2$ & $\Delta$RV$_2$\\
\text{[d]} & [km\,s$^{-1}$] & [km\,s$^{-1}$] & [km\,s$^{-1}$] & [km\,s$^{-1}$] \\

\hline
8644.88861 & +100.1 & $\pm$3.5 & $-$66.0 & $\pm$5.0\\
8644.90609 & +106.5 & $\pm$3.5 & $-$76.5 & $\pm$4.3\\
8645.89599 & +0.6 & $\pm$2.5 & $-$2.1 & $\pm$4.7\\
9007.92584 & $-$68.9 & $\pm$10.2 & +125.5 & $\pm$16.9\\
9007.93176 & $-$72.7 & $\pm$11.8 & +105.6 & $\pm$8.5\\
9008.91465 & +8.2 & $\pm$11.7 & $-$33.2 & $\pm$23.8\\
9008.92058 & $-$3.6 & $\pm$5.8 & $-$13.8 & $\pm$17.9\\
9008.92746 & $-$14.8 & $\pm$15.1 & +0.6 & $\pm$34.8\\
9008.93338 & $-$7.1 & $\pm$4.4 & $-$8.9 & $\pm$14.8\\
9008.97623 & +53.1 & $\pm$20.4 & $-$57.8 & $\pm$33.3\\
9008.98216 & +77.3 & $\pm$21.3 & $-$43.5 & $\pm$35.4\\
9010.88005 & $-$0.0 & $\pm$28.9 & +33.1 & $\pm$30.1\\
9010.88598 & $-$82.4 & $\pm$18.7 & +80.3 & $\pm$27.3\\
9012.96409 & +61.0 & $\pm$20.7 & $-$90.3 & $\pm$14.9\\
9012.97117 & +56.1 & $\pm$14.7 & $-$93.2 & $\pm$34.6\\
10034.98744 & +109.1 & $\pm$37.4 & $-$106.5 & $\pm$30.1\\
10034.99916 & +66.1 & $\pm$32.7 & $-$92.4 & $\pm$55.6\\
10035.01088 & +110.1 & $\pm$16.2 & $-$108.6 & $\pm$45.3\\
10035.09578 & $-$13.8 & $\pm$9.9 & $-$18.4 & $\pm$32.9\\
10035.10749 & +1.3 & $\pm$6.6 & $-$6.5 & $\pm$37.0\\
10035.99829 & $-$86.5 & $\pm$30.6 & +39.8 & $\pm$34.7\\
10036.01000 & $-$132.0 & $\pm$28.0 & +135.2 & $\pm$24.4\\
10036.02172 & $-$156.1 & $\pm$13.3 & +75.9 & $\pm$33.5\\
10036.03412 & $-$85.9 & $\pm$21.1 & +128.6 & $\pm$34.1\\
10425.89014 & +62.5 & $\pm$13.1 & $-$64.7 & $\pm$20.6\\
10425.91157 & +74.1 & $\pm$15.8 & $-$114.5 & $\pm$25.8\\
10425.93311 & +85.2 & $\pm$6.4 & $-$115.3 & $\pm$14.1\\
10426.88854 & +43.9 & $\pm$15.5 & $-$47.2 & $\pm$22.1\\
10426.91003 & +29.7 & $\pm$11.2 & $-$25.6 & $\pm$18.5\\
10426.93149 & +3.9 & $\pm$7.3 & +4.9 & $\pm$11.4\\
10426.95653 & $-$16.1 & $\pm$9.1 & +15.2 & $\pm$19.3\\
10426.97801 & $-$11.8 & $\pm$13.0 & +61.1 & $\pm$17.0\\
10426.99960 & $-$42.3 & $\pm$11.2 & +84.7 & $\pm$12.8\\
10429.95934 & $-$15.9 & $\pm$19.7 & +15.5 & $\pm$19.9\\
10429.98088 & $-$40.1 & $\pm$22.5 & +48.5 & $\pm$15.5\\
10430.00272 & $-$40.0 & $\pm$15.5 & +81.4 & $\pm$14.7\\
10430.87167 & $-$76.0 & $\pm$11.9 & +137.4 & $\pm$8.3\\
10430.89760 & $-$84.6 & $\pm$8.5 & +111.4 & $\pm$9.5\\
10430.91913 & $-$74.2 & $\pm$18.2 & +102.6 & $\pm$14.1\\
10431.91069 & +79.9 & $\pm$21.2 & $-$111.5 & $\pm$17.4\\
10431.93222 & +59.4 & $\pm$20.6 & $-$127.9 & $\pm$11.0\\
10431.94791 & +141.4 & $\pm$35.9 & $-$119.8 & $\pm$13.8\\
10431.97491 & +100.0 & $\pm$18.3 & $-$151.3 & $\pm$20.7\\
10431.99643 & +105.8 & $\pm$10.4 & $-$144.0 & $\pm$16.1\\
\end{tabular}
\label{tab:AppendixRVsWDJ114446}
\end{table}

\begin{table}
\centering
\caption{RVs for the target WDJ141625.94+311600.55}
\begin{tabular}{r|r|r|r|r} \\
HJD$-$2450000 & $\Delta$RV$_1$ & RV$_1$ & RV$_2$ & $\Delta$RV$_2$\\
\text{[d]} & [km\,s$^{-1}$] & [km\,s$^{-1}$] & [km\,s$^{-1}$] & [km\,s$^{-1}$] \\

\hline
8590.01924 & $-$41.3 & $\pm$14.8 & +150.1 & $\pm$10.0\\
8590.03557 & $-$59.0 & $\pm$20.8 & +113.8 & $\pm$12.6\\
8590.96439 & +107.8 & $\pm$9.9 & $-$36.7 & $\pm$6.7\\
8590.98072 & +112.9 & $\pm$9.2 & $-$24.2 & $\pm$8.6\\
8671.89349 & +17.6 & $\pm$10.6 & +44.8 & $\pm$8.8\\
10342.17825 & +44.4 & $\pm$13.4 & +2.1 & $\pm$10.4\\
10342.20083 & +100.6 & $\pm$20.9 & +23.4 & $\pm$13.9\\
10342.22315 & +131.1 & $\pm$16.0 & $-$1.0 & $\pm$12.0\\
10342.26459 & +108.9 & $\pm$9.8 & $-$47.7 & $\pm$7.0\\
10429.87151 & +37.8 & $\pm$14.3 & +33.7 & $\pm$7.6\\
10429.89300 & +20.2 & $\pm$11.9 & +20.9 & $\pm$10.6\\
10429.91455 & +21.1 & $\pm$22.4 & +20.6 & $\pm$15.3\\
10429.93606 & +38.3 & $\pm$16.3 & +47.6 & $\pm$10.5\\
10430.02894 & $-$80.5 & $\pm$16.6 & +93.8 & $\pm$16.5\\
10430.05352 & $-$113.0 & $\pm$23.0 & +116.6 & $\pm$11.6\\
10430.07766 & $-$95.7 & $\pm$42.9 & +124.8 & $\pm$14.4\\
10430.09926 & $-$108.4 & $\pm$12.3 & +55.1 & $\pm$33.3\\
10430.12075 & $-$110.6 & $\pm$12.2 & +135.3 & $\pm$16.1\\
10430.94581 & $-$55.2 & $\pm$26.4 & +77.0 & $\pm$21.3\\
10430.96679 & $-$84.5 & $\pm$19.9 & +97.9 & $\pm$20.7\\
10430.98832 & $-$58.0 & $\pm$21.9 & +56.0 & $\pm$23.8\\
10432.98031 & +14.5 & $\pm$16.0 & +18.2 & $\pm$8.6\\
10433.00185 & +28.5 & $\pm$9.3 & +28.6 & $\pm$7.5\\
10699.23763 & $-$95.4 & $\pm$11.6 & +112.9 & $\pm$10.3\\
10700.21245 & +94.1 & $\pm$14.9 & +10.3 & $\pm$12.8\\
10700.23370 & +89.0 & $\pm$13.8 & $-$6.2 & $\pm$18.1\\
10701.22036 & +121.7 & $\pm$7.2 & $-$16.9 & $\pm$12.8\\
10701.24160 & +114.8 & $\pm$6.2 & $-$23.9 & $\pm$9.0\\
10701.26283 & +84.1 & $\pm$18.2 & +13.8 & $\pm$22.3\\
10702.15219 & $-$50.1 & $\pm$7.7 & +86.4 & $\pm$8.0\\
10702.17342 & $-$54.1 & $\pm$13.3 & +88.1 & $\pm$6.6\\
10702.22994 & $-$93.7 & $\pm$13.2 & +113.5 & $\pm$12.6\\
10703.12908 & $-$70.0 & $\pm$5.3 & +96.9 & $\pm$9.9\\
\end{tabular}
\label{tab:AppendixRVsWDJ141625}
\end{table}

\begin{table}
\centering
\caption{RVs for the target WDJ151109.90+404801.18. Timestamps with asterisks next to the numbers indicate cases where it is unclear if the star 1/2 assignments are correct.}
\begin{tabular}{r|r|r|r|r} \\
HJD$-$2450000 & $\Delta$RV$_1$ & RV$_1$ & RV$_2$ & $\Delta$RV$_2$\\
\text{[d]} & [km\,s$^{-1}$] & [km\,s$^{-1}$] & [km\,s$^{-1}$] & [km\,s$^{-1}$] \\

\hline
8645.04282 & +80.1 & $\pm$6.2 & $-$61.0 & $\pm$5.9\\
8645.05684 & +70.8 & $\pm$2.8 & $-$64.7 & $\pm$2.5\\
$*$10035.13492 & +27.2 & $\pm$17.6 & $-$121.7 & $\pm$22.1\\
$*$10035.15670 & +33.1 & $\pm$20.1 & $-$81.8 & $\pm$13.6\\
$*$10035.21695 & +36.6 & $\pm$19.2 & $-$46.8 & $\pm$13.2\\
$*$10035.23867 & +67.4 & $\pm$12.4 & $-$29.5 & $\pm$13.6\\
$*$10036.18014 & +26.2 & $\pm$25.8 & $-$44.6 & $\pm$34.0\\
10425.95913 & $-$97.0 & $\pm$11.6 & +72.5 & $\pm$8.5\\
10425.98056 & $-$68.1 & $\pm$11.4 & +80.2 & $\pm$7.4\\
10426.00206 & $-$92.3 & $\pm$7.0 & +77.8 & $\pm$6.3\\
10427.02513 & $-$11.4 & $\pm$16.7 & +95.1 & $\pm$12.2\\
10427.04667 & $-$7.1 & $\pm$20.1 & +52.1 & $\pm$62.6\\
10431.01307 & +36.4 & $\pm$9.9 & $-$55.5 & $\pm$15.9\\
10431.03460 & $-$8.7 & $\pm$12.8 & $-$100.1 & $\pm$11.5\\
10431.05606 & +10.3 & $\pm$14.5 & $-$120.0 & $\pm$17.8\\
10431.07759 & +40.3 & $\pm$8.9 & $-$99.7 & $\pm$11.8\\
10431.19518 & +42.7 & $\pm$9.3 & $-$105.4 & $\pm$9.0\\
10431.21661 & +44.6 & $\pm$8.9 & $-$84.0 & $\pm$6.9\\
10432.02148 & +71.9 & $\pm$12.7 & $-$45.9 & $\pm$7.6\\
10432.04296 & +54.3 & $\pm$10.8 & $-$71.0 & $\pm$10.0\\
10432.06449 & +62.0 & $\pm$9.4 & $-$81.5 & $\pm$14.4\\
10433.98190 & +37.2 & $\pm$15.7 & $-$77.8 & $\pm$53.0\\
10434.00343 & +72.9 & $\pm$9.3 & $-$55.9 & $\pm$9.6\\
\end{tabular}
\label{tab:AppendixRVsWDJ151109}
\end{table}

\begin{table}
\centering
\caption{RVs for the target WDJ153615.83+501350.98}
\begin{tabular}{r|r|r|r|r} \\
HJD$-$2450000 & $\Delta$RV$_1$ & RV$_1$ & RV$_2$ & $\Delta$RV$_2$\\
\text{[d]} & [km\,s$^{-1}$] & [km\,s$^{-1}$] & [km\,s$^{-1}$] & [km\,s$^{-1}$] \\

\hline
8645.07514 & +77.5 & $\pm$4.2 & +13.4 & $\pm$7.6\\
8645.08915 & +86.8 & $\pm$5.8 & $-$26.5 & $\pm$9.1\\
\end{tabular}
\label{tab:AppendixRVsWDJ153615}
\end{table}

\begin{table}
\centering
\caption{RVs for the target WDJ160822.19+420543.44}
\begin{tabular}{r|r|r|r|r} \\
HJD$-$2450000 & $\Delta$RV$_1$ & RV$_1$ & RV$_2$ & $\Delta$RV$_2$\\
\text{[d]} & [km\,s$^{-1}$] & [km\,s$^{-1}$] & [km\,s$^{-1}$] & [km\,s$^{-1}$] \\

\hline
8646.08533 & $-$117.4 & $\pm$2.5 & +42.0 & $\pm$2.4\\
8646.09240 & $-$110.0 & $\pm$2.5 & +42.9 & $\pm$3.2\\
8646.09946 & $-$108.2 & $\pm$2.4 & +34.3 & $\pm$3.2\\
8646.11491 & $-$96.0 & $\pm$1.5 & +30.4 & $\pm$1.7\\
8646.12745 & $-$82.0 & $\pm$1.8 & +23.5 & $\pm$2.3\\
9008.01298 & +37.7 & $\pm$7.2 & $-$68.7 & $\pm$7.9\\
9008.02006 & +43.8 & $\pm$6.7 & $-$79.0 & $\pm$7.9\\
9008.94875 & +88.0 & $\pm$7.4 & $-$101.9 & $\pm$7.4\\
9008.95467 & +90.1 & $\pm$6.1 & $-$106.8 & $\pm$7.8\\
9008.96157 & +98.2 & $\pm$7.2 & $-$102.8 & $\pm$8.8\\
9008.96749 & +103.6 & $\pm$7.5 & $-$128.3 & $\pm$10.3\\
9010.94928 & $-$113.1 & $\pm$9.7 & +36.9 & $\pm$8.0\\
9010.95521 & $-$103.3 & $\pm$11.4 & +15.8 & $\pm$10.7\\
9012.04518 & $-$118.4 & $\pm$7.5 & +33.5 & $\pm$9.4\\
9012.05226 & $-$101.3 & $\pm$5.1 & +36.9 & $\pm$3.5\\
9012.12913 & $-$22.9 & $\pm$6.9 & $-$23.8 & $\pm$7.9\\
9012.13621 & $-$20.9 & $\pm$10.8 & $-$24.7 & $\pm$11.2\\
10035.17543 & $-$132.5 & $\pm$17.9 & +39.6 & $\pm$24.5\\
10035.18020 & $-$96.6 & $\pm$15.5 & +45.6 & $\pm$10.6\\
10035.18497 & $-$108.1 & $\pm$24.3 & +51.6 & $\pm$26.2\\
10035.18974 & $-$166.9 & $\pm$10.9 & +49.8 & $\pm$11.6\\
10035.25365 & $-$92.1 & $\pm$27.3 & +14.2 & $\pm$16.9\\
10036.06601 & $-$101.5 & $\pm$13.7 & +16.9 & $\pm$15.3\\
10036.07282 & $-$92.0 & $\pm$34.1 & +30.9 & $\pm$31.1\\
10036.07922 & $-$93.1 & $\pm$12.0 & +27.9 & $\pm$11.8\\
10036.08561 & $-$53.6 & $\pm$7.5 & +40.8 & $\pm$5.8\\
10036.21760 & +31.2 & $\pm$39.1 & $-$12.8 & $\pm$47.8\\
10036.22399 & +42.9 & $\pm$15.0 & $-$72.1 & $\pm$8.2\\
10036.23038 & +61.8 & $\pm$10.0 & $-$49.7 & $\pm$16.1\\
10036.23810 & +52.8 & $\pm$15.2 & $-$98.6 & $\pm$15.5\\
\end{tabular}
\label{tab:AppendixRVsWDJ160822}
\end{table}

\begin{table}
\centering
\caption{RVs for the target WDJ163441.85+173634.09}
\begin{tabular}{r|r|r|r|r} \\
HJD$-$2450000 & $\Delta$RV$_1$ & RV$_1$ & RV$_2$ & $\Delta$RV$_2$\\
\text{[d]} & [km\,s$^{-1}$] & [km\,s$^{-1}$] & [km\,s$^{-1}$] & [km\,s$^{-1}$] \\

\hline
8645.10856 & +88.7 & $\pm$2.5 & $-$33.3 & $\pm$4.4\\
8645.11563 & +88.9 & $\pm$2.2 & $-$33.5 & $\pm$3.8\\
8645.12270 & +89.4 & $\pm$2.1 & $-$31.4 & $\pm$3.3\\
9007.98904 & $-$28.8 & $\pm$7.6 & +60.2 & $\pm$7.9\\
9007.99612 & $-$35.4 & $\pm$5.3 & +71.5 & $\pm$7.9\\
9009.14135 & +39.9 & $\pm$5.9 & +13.2 & $\pm$10.7\\
9009.14727 & +61.8 & $\pm$4.2 & +2.5 & $\pm$6.9\\
9010.96912 & +73.6 & $\pm$10.7 & $-$28.8 & $\pm$18.1\\
9010.97505 & +68.2 & $\pm$8.6 & $-$44.9 & $\pm$15.1\\
9010.98196 & +93.8 & $\pm$8.2 & $-$26.1 & $\pm$10.6\\
9010.98788 & +62.9 & $\pm$13.9 & $-$20.4 & $\pm$16.6\\
9012.01278 & $-$54.3 & $\pm$4.8 & +69.9 & $\pm$6.8\\
9012.01986 & $-$53.3 & $\pm$5.7 & +71.8 & $\pm$8.8\\
9012.02791 & $-$51.6 & $\pm$3.9 & +78.0 & $\pm$5.3\\
9012.03499 & $-$44.8 & $\pm$4.9 & +74.2 & $\pm$7.4\\
\end{tabular}
\label{tab:AppendixRVsWDJ163441}
\end{table}

\begin{table}
\centering
\caption{RVs for the target WDJ182606.04+482911.30}
\begin{tabular}{r|r|r|r|r} \\
HJD$-$2450000 & $\Delta$RV$_1$ & RV$_1$ & RV$_2$ & $\Delta$RV$_2$\\
\text{[d]} & [km\,s$^{-1}$] & [km\,s$^{-1}$] & [km\,s$^{-1}$] & [km\,s$^{-1}$] \\

\hline
8359.92351 & +6.2 & $\pm$6.9 & $-$95.7 & $\pm$8.8\\
8367.02429 & +36.8 & $\pm$5.5 & $-$120.6 & $\pm$13.7\\
8725.89196 & $-$70.7 & $\pm$15.7 & +87.3 & $\pm$30.3\\
8725.91296 & $-$61.2 & $\pm$22.0 & +83.5 & $\pm$25.7\\
8725.93608 & $-$101.4 & $\pm$17.2 & +83.7 & $\pm$30.4\\
8725.95705 & $-$95.8 & $\pm$10.5 & +72.8 & $\pm$22.8\\
8728.98501 & $-$16.4 & $\pm$9.2 & $-$6.4 & $\pm$28.7\\
8729.00601 & $-$12.3 & $\pm$16.9 & $-$65.9 & $\pm$42.6\\
8729.02936 & $-$56.8 & $\pm$30.1 & +31.1 & $\pm$20.2\\
8729.05033 & $-$40.1 & $\pm$22.4 & +45.0 & $\pm$20.4\\
8733.90809 & $-$5.0 & $\pm$7.1 & $-$12.0 & $\pm$37.4\\
8733.92906 & $-$36.7 & $\pm$14.0 & $-$67.9 & $\pm$31.2\\
10431.09891 & +31.1 & $\pm$14.5 & $-$100.3 & $\pm$21.3\\
10431.12044 & +20.7 & $\pm$30.6 & $-$71.5 & $\pm$39.4\\
10431.14193 & +11.6 & $\pm$22.3 & $-$100.7 & $\pm$25.2\\
10431.16346 & +37.2 & $\pm$12.1 & $-$97.0 & $\pm$16.3\\
10432.12048 & $-$15.9 & $\pm$21.7 & $-$40.0 & $\pm$18.7\\
10432.16354 & $-$11.7 & $\pm$17.1 & $-$73.5 & $\pm$29.5\\
10432.18534 & +2.9 & $\pm$17.6 & $-$46.5 & $\pm$27.4\\
10433.02499 & $-$32.4 & $\pm$13.7 & +29.4 & $\pm$37.3\\
10433.04651 & $-$39.9 & $\pm$16.8 & $-$5.0 & $\pm$25.6\\
10434.02792 & +25.5 & $\pm$20.1 & $-$75.8 & $\pm$56.3\\
10434.04937 & +68.8 & $\pm$11.9 & $-$108.7 & $\pm$14.9\\
10434.19949 & +36.4 & $\pm$31.7 & $-$123.1 & $\pm$14.8\\
\end{tabular}
\label{tab:AppendixRVsWDJ182606}
\end{table}

\begin{table}
\centering
\caption{RVs for the target WDJ183442.33-170028.00}
\begin{tabular}{r|r|r|r|r} \\
HJD$-$2450000 & $\Delta$RV$_1$ & RV$_1$ & RV$_2$ & $\Delta$RV$_2$\\
\text{[d]} & [km\,s$^{-1}$] & [km\,s$^{-1}$] & [km\,s$^{-1}$] & [km\,s$^{-1}$] \\

\hline
8364.89558 & $-$177.2 & $\pm$3.3 & +119.4 & $\pm$11.2\\
8366.91659 & $-$133.7 & $\pm$4.2 & +92.7 & $\pm$12.2\\
8366.93753 & $-$103.6 & $\pm$3.6 & +67.5 & $\pm$17.5\\
8671.00733 & $-$169.2 & $\pm$4.6 & +118.7 & $\pm$14.4\\
8671.02375 & $-$154.4 & $\pm$3.1 & +117.0 & $\pm$13.0\\
8672.05606 & $-$86.1 & $\pm$5.6 & +103.6 & $\pm$6.9\\
9010.05790 & $-$57.4 & $\pm$17.2 & +117.6 & $\pm$43.9\\
9010.08582 & $-$24.4 & $\pm$41.2 & +87.7 & $\pm$54.8\\
9012.18969 & $-$114.3 & $\pm$13.0 & +149.1 & $\pm$23.8\\
9014.07330 & $-$81.0 & $\pm$11.9 & +113.3 & $\pm$13.3\\
9014.10122 & $-$130.6 & $\pm$7.4 & +107.7 & $\pm$16.7\\
9014.13012 & $-$170.6 & $\pm$10.2 & +168.1 & $\pm$23.9\\
9014.15804 & $-$137.8 & $\pm$15.3 & +131.5 & $\pm$14.4\\
10432.21371 & $-$45.6 & $\pm$35.4 & +96.6 & $\pm$69.6\\
10434.16487 & +52.6 & $\pm$17.5 & $-$34.8 & $\pm$42.0\\
10475.10559 & +159.9 & $\pm$25.2 & $-$129.7 & $\pm$42.9\\
10475.12747 & +123.9 & $\pm$26.8 & $-$91.0 & $\pm$58.0\\
\end{tabular}
\label{tab:AppendixRVsWDJ183442}
\end{table}

\begin{table}
\centering
\caption{RVs for the target WDJ212935.23+001332.26}
\begin{tabular}{r|r|r|r|r} \\
HJD$-$2450000 & $\Delta$RV$_1$ & RV$_1$ & RV$_2$ & $\Delta$RV$_2$\\
\text{[d]} & [km\,s$^{-1}$] & [km\,s$^{-1}$] & [km\,s$^{-1}$] & [km\,s$^{-1}$] \\

\hline
8359.07483 & +82.0 & $\pm$4.4 & $-$21.2 & $\pm$4.8\\
8367.07114 & +134.5 & $\pm$4.5 & $-$54.4 & $\pm$6.3\\
8723.90031 & $-$23.1 & $\pm$16.0 & +63.7 & $\pm$18.3\\
8723.92587 & +3.1 & $\pm$24.3 & +52.2 & $\pm$57.1\\
8723.94684 & +29.7 & $\pm$11.7 & +31.4 & $\pm$16.5\\
8725.10960 & $-$67.2 & $\pm$10.0 & +110.6 & $\pm$14.0\\
8725.13057 & $-$49.6 & $\pm$7.4 & +96.6 & $\pm$11.7\\
8725.98813 & +139.5 & $\pm$6.7 & $-$45.9 & $\pm$12.2\\
8726.00910 & +140.3 & $\pm$7.0 & $-$65.5 & $\pm$15.0\\
8726.03216 & +119.0 & $\pm$7.3 & $-$46.8 & $\pm$9.0\\
8726.05313 & +119.5 & $\pm$8.6 & $-$46.0 & $\pm$22.6\\
8726.07613 & +95.9 & $\pm$6.9 & $-$33.7 & $\pm$8.7\\
8726.09711 & +73.4 & $\pm$7.5 & $-$6.8 & $\pm$14.6\\
8730.11932 & $-$37.5 & $\pm$8.7 & +134.8 & $\pm$18.3\\
\end{tabular}
\label{tab:AppendixRVsWDJ212935}
\end{table}

\begin{table}
\centering
\caption{RVs for the target WDJ231404.30+552814.11}
\begin{tabular}{r|r|r|r|r} \\
HJD$-$2450000 & $\Delta$RV$_1$ & RV$_1$ & RV$_2$ & $\Delta$RV$_2$\\
\text{[d]} & [km\,s$^{-1}$] & [km\,s$^{-1}$] & [km\,s$^{-1}$] & [km\,s$^{-1}$] \\

\hline
8360.06566 & $-$58.3 & $\pm$7.6 & +42.5 & $\pm$4.9\\
8362.16173 & +73.8 & $\pm$5.9 & $-$53.5 & $\pm$4.8\\
8362.17920 & +59.3 & $\pm$5.6 & $-$50.4 & $\pm$4.4\\
8725.15411 & $-$37.8 & $\pm$13.8 & +28.2 & $\pm$10.7\\
8725.17508 & $-$27.5 & $\pm$15.5 & +22.3 & $\pm$12.6\\
8725.19819 & $-$5.0 & $\pm$21.8 & +19.5 & $\pm$16.6\\
8725.21916 & $-$2.2 & $\pm$16.8 & +4.0 & $\pm$12.8\\
8725.24444 & +25.9 & $\pm$13.7 & +29.5 & $\pm$11.7\\
8727.06440 & $-$34.1 & $\pm$25.0 & +36.5 & $\pm$17.1\\
8727.10831 & $-$5.3 & $\pm$31.2 & +10.5 & $\pm$28.8\\
8727.12928 & $-$3.5 & $\pm$33.8 & $-$9.4 & $\pm$27.7\\
8727.15169 & $-$47.0 & $\pm$7.7 & +32.0 & $\pm$10.1\\
8727.88956 & $-$60.6 & $\pm$19.5 & +82.3 & $\pm$12.5\\
8727.91055 & $-$38.1 & $\pm$10.0 & +65.2 & $\pm$8.5\\
8727.93548 & $-$44.7 & $\pm$8.9 & +78.9 & $\pm$10.7\\
8727.95646 & $-$35.4 & $\pm$5.8 & +96.1 & $\pm$8.2\\
8733.11161 & +33.7 & $\pm$15.1 & $-$29.6 & $\pm$8.8\\
9794.51103 & +72.6 & $\pm$15.7 & $-$48.5 & $\pm$13.5\\
9794.62542 & +76.2 & $\pm$12.2 & $-$62.4 & $\pm$10.6\\
9795.38290 & $-$38.5 & $\pm$14.4 & +47.0 & $\pm$11.1\\
9795.46051 & $-$70.8 & $\pm$18.1 & +53.6 & $\pm$12.1\\
9797.49635 & +58.7 & $\pm$13.0 & $-$72.2 & $\pm$12.2\\
9797.59125 & +57.2 & $\pm$14.0 & $-$60.9 & $\pm$10.0\\
9799.46266 & $-$18.8 & $\pm$28.0 & $-$5.9 & $\pm$31.2\\
9799.61133 & $-$15.1 & $\pm$20.5 & $-$17.5 & $\pm$15.5\\
9800.56944 & +40.8 & $\pm$15.4 & $-$76.3 & $\pm$12.6\\
10817.15999 & $-$94.3 & $\pm$12.7 & +62.0 & $\pm$10.4\\
10817.21082 & $-$80.3 & $\pm$12.0 & +58.4 & $\pm$7.2\\
10820.14396 & $-$64.5 & $\pm$7.9 & +72.9 & $\pm$5.4\\
10820.20666 & $-$68.2 & $\pm$7.6 & +80.1 & $\pm$5.2\\
10826.21066 & $-$61.5 & $\pm$6.6 & +68.3 & $\pm$4.3\\
\end{tabular}
\label{tab:AppendixRVsWDJ231404}
\end{table}

\section{RVs of double lined systems with period aliases}
\label{appendix:RVsAliases}
Found in Tables~\ref{tab:AppendixRVsWDJ000319}--\ref{tab:AppendixRVsWDJ180115} are the observed RVs of every source that have multiple viable period aliases.
\begin{table}
\centering
\caption{RVs for the target WDJ000319.54+022623.28}
\begin{tabular}{r|r|r|r|r} \\
HJD-2450000 & $\Delta$RV$_1$ & RV$_1$ & RV$_2$ & $\Delta$RV$_2$\\
\text{[d]} & [km\,s$^{-1}$] & [km\,s$^{-1}$] & [km\,s$^{-1}$] & [km\,s$^{-1}$] \\

\hline
8671.21502 & +56.0 & $\pm$5.9 & +19.8 & $\pm$16.1\\
8672.21313 & +99.4 & $\pm$5.0 & $-$51.9 & $\pm$12.1\\
8672.22725 & +98.3 & $\pm$4.8 & $-$79.8 & $\pm$12.2\\
8727.18346 & $-$10.7 & $\pm$10.9 & +146.0 & $\pm$24.9\\
8727.20443 & $-$18.9 & $\pm$6.8 & +113.9 & $\pm$16.7\\
8727.22709 & $-$3.5 & $\pm$5.0 & +172.2 & $\pm$19.4\\
8727.24806 & $-$9.2 & $\pm$16.4 & +144.6 & $\pm$29.4\\
8727.98587 & +94.4 & $\pm$12.7 & $-$12.4 & $\pm$36.3\\
8728.00687 & +119.7 & $\pm$12.7 & $-$66.4 & $\pm$35.0\\
8728.02931 & +131.7 & $\pm$22.8 & $-$45.2 & $\pm$45.3\\
8728.05028 & +116.1 & $\pm$13.5 & $-$47.5 & $\pm$30.2\\
8728.07276 & +110.4 & $\pm$14.1 & $-$116.2 & $\pm$31.2\\
8728.09373 & +104.6 & $\pm$10.1 & $-$120.0 & $\pm$13.0\\
8729.13332 & +38.8 & $\pm$13.8 & $-$14.5 & $\pm$35.0\\
8729.15429 & +63.0 & $\pm$8.3 & $-$22.9 & $\pm$32.9\\
8729.17799 & +67.1 & $\pm$20.6 & $-$44.8 & $\pm$33.5\\
8731.18880 & +100.5 & $\pm$6.9 & +20.0 & $\pm$48.1\\
\end{tabular}
\label{tab:AppendixRVsWDJ000319}
\end{table}

\begin{table}
\centering
\caption{RVs for the target WDJ080856.79+461300.08}
\begin{tabular}{r|r|r|r|r} \\
HJD-2450000 & $\Delta$RV$_1$ & RV$_1$ & RV$_2$ & $\Delta$RV$_2$\\
\text{[d]} & [km\,s$^{-1}$] & [km\,s$^{-1}$] & [km\,s$^{-1}$] & [km\,s$^{-1}$] \\

\hline
8527.96071 & +97.8 & $\pm$22.9 & $-$24.3 & $\pm$35.9\\
8534.12860 & $-$16.9 & $\pm$8.1 & +111.7 & $\pm$11.5\\
8534.15398 & $-$16.9 & $\pm$8.7 & +88.1 & $\pm$15.9\\
10432.86430 & +63.0 & $\pm$13.1 & $-$71.4 & $\pm$33.5\\
10432.88574 & +76.5 & $\pm$15.7 & $-$34.8 & $\pm$54.0\\
10432.90727 & +89.9 & $\pm$18.3 & $-$10.3 & $\pm$27.2\\
10432.92871 & +112.1 & $\pm$13.7 & $-$0.2 & $\pm$28.6\\
10432.95016 & +103.8 & $\pm$14.3 & $-$15.9 & $\pm$19.6\\
10433.86435 & +79.1 & $\pm$16.6 & $-$25.3 & $\pm$46.1\\
10433.88588 & +120.9 & $\pm$12.3 & $-$75.4 & $\pm$20.9\\
10433.90923 & +127.9 & $\pm$12.4 & $-$38.4 & $\pm$26.1\\
10433.93076 & +116.6 & $\pm$22.9 & $-$54.4 & $\pm$45.8\\
10433.95221 & +113.4 & $\pm$18.6 & $-$51.2 & $\pm$36.4\\
10461.90284 & $-$29.2 & $\pm$22.1 & +139.2 & $\pm$30.2\\
10461.92407 & $-$10.9 & $\pm$7.8 & +133.1 & $\pm$14.8\\
10752.97090 & +89.5 & $\pm$12.8 & $-$13.2 & $\pm$17.4\\
10753.03265 & +66.3 & $\pm$15.6 & $-$17.3 & $\pm$25.2\\
\end{tabular}
\label{tab:AppendixRVsWDJ080856}
\end{table}

\begin{table}
\centering
\caption{RVs for the target WDJ170120.99$-$191527.57}
\begin{tabular}{r|r|r|r|r} \\
HJD-2450000 & $\Delta$RV$_1$ & RV$_1$ & RV$_2$ & $\Delta$RV$_2$\\
\text{[d]} & [km\,s$^{-1}$] & [km\,s$^{-1}$] & [km\,s$^{-1}$] & [km\,s$^{-1}$] \\

\hline
8589.23520 & $-$140.0 & $\pm$12.4 & +102.3 & $\pm$11.4\\
8591.15233 & +58.1 & $\pm$6.9 & $-$83.8 & $\pm$4.6\\
8591.16866 & +116.4 & $\pm$9.2 & $-$137.4 & $\pm$5.6\\
10430.14734 & $-$160.3 & $\pm$22.4 & +98.1 & $\pm$14.4\\
10430.16926 & $-$139.4 & $\pm$18.4 & +80.5 & $\pm$10.7\\
10430.19189 & $-$150.8 & $\pm$19.4 & +48.9 & $\pm$18.7\\
10433.07839 & $-$117.1 & $\pm$16.2 & +80.8 & $\pm$16.3\\
10433.10133 & $-$54.8 & $\pm$15.5 & +20.3 & $\pm$14.2\\
10433.12283 & +44.5 & $\pm$17.0 & $-$60.3 & $\pm$14.1\\
10433.14441 & +91.5 & $\pm$21.1 & $-$129.7 & $\pm$34.9\\
10433.16594 & +120.0 & $\pm$10.3 & $-$140.9 & $\pm$11.6\\
10434.08238 & $-$148.6 & $\pm$12.8 & +78.1 & $\pm$11.8\\
10434.09934 & $-$121.5 & $\pm$22.7 & +93.1 & $\pm$13.9\\
10434.11718 & $-$98.7 & $\pm$41.4 & +76.3 & $\pm$11.3\\
10434.13872 & $-$31.2 & $\pm$36.8 & $-$7.4 & $\pm$25.4\\
10460.06880 & $-$153.4 & $\pm$9.9 & +112.9 & $\pm$8.2\\
10460.07816 & $-$153.5 & $\pm$14.1 & +118.0 & $\pm$10.0\\
10460.08723 & $-$113.1 & $\pm$11.5 & +85.7 & $\pm$8.3\\
10460.09652 & $-$112.3 & $\pm$16.1 & +59.0 & $\pm$12.3\\
10460.10560 & $-$82.8 & $\pm$13.5 & +57.2 & $\pm$13.0\\
10460.11494 & $-$55.9 & $\pm$11.8 & +6.9 & $\pm$13.9\\
10460.12403 & $-$19.3 & $\pm$19.2 & $-$16.4 & $\pm$13.4\\
10460.13729 & +41.9 & $\pm$15.9 & $-$47.0 & $\pm$27.3\\
10460.14663 & +80.3 & $\pm$14.8 & $-$95.6 & $\pm$12.7\\
10460.15571 & +103.4 & $\pm$11.6 & $-$112.4 & $\pm$9.2\\
10463.05503 & +120.4 & $\pm$17.0 & $-$135.2 & $\pm$11.8\\
10463.06403 & +99.8 & $\pm$18.1 & $-$132.9 & $\pm$13.5\\
10463.18043 & $-$142.4 & $\pm$14.6 & +103.4 & $\pm$12.3\\
10502.10991 & $-$101.8 & $\pm$3.8 & +57.0 & $\pm$4.3\\
10502.12407 & $-$120.8 & $\pm$4.1 & +72.7 & $\pm$3.7\\
\end{tabular}
\label{tab:AppendixRVsWDJ170120}
\end{table}

\begin{table}
\centering
\caption{RVs for the target WDJ180115.37+721848.76}
\begin{tabular}{r|r|r|r|r} \\
HJD-2450000 & $\Delta$RV$_1$ & RV$_1$ & RV$_2$ & $\Delta$RV$_2$\\
\text{[d]} & [km\,s$^{-1}$] & [km\,s$^{-1}$] & [km\,s$^{-1}$] & [km\,s$^{-1}$] \\

\hline
8359.88252 & $-$94.0 & $\pm$3.9 & +106.5 & $\pm$17.6\\
8363.86443 & +53.6 & $\pm$3.7 & $-$71.8 & $\pm$8.9\\
8363.88191 & +61.9 & $\pm$3.3 & $-$79.0 & $\pm$7.5\\
9007.95255 & $-$0.0 & $\pm$7.8 & +1.1 & $\pm$36.2\\
9007.96657 & $-$11.3 & $\pm$13.9 & +75.9 & $\pm$25.7\\
9009.95799 & +109.3 & $\pm$23.7 & $-$124.7 & $\pm$55.8\\
9009.97549 & +116.7 & $\pm$18.8 & $-$127.4 & $\pm$46.0\\
9012.92967 & $-$67.7 & $\pm$12.4 & +46.9 & $\pm$50.8\\
9012.94647 & $-$77.0 & $\pm$12.9 & +46.8 & $\pm$31.7\\
10700.26108 & $-$91.9 & $\pm$7.8 & +100.6 & $\pm$19.0\\
10700.28582 & $-$90.6 & $\pm$5.7 & +139.5 & $\pm$11.8\\
10701.28668 & $-$45.8 & $\pm$10.0 & +64.8 & $\pm$42.6\\
10702.27395 & +65.3 & $\pm$6.8 & $-$92.5 & $\pm$45.3\\
10702.29171 & +57.9 & $\pm$10.0 & $-$130.2 & $\pm$43.7\\
10703.26520 & +108.7 & $\pm$4.2 & $-$101.5 & $\pm$19.8\\
10703.28644 & +112.0 & $\pm$5.0 & $-$104.9 & $\pm$18.0\\
\end{tabular}
\label{tab:AppendixRVsWDJ180115}
\end{table}

\section{Radial velocities of single-lined sources}
\label{appendix:RVsSB1}
Found in Table~\ref{tab:DBLsinglelined} are the RVs extracted for all single-lined sources observed in the DBL survey where hydrogen Balmer absorption lines are apparent.

\begin{table}
\centering
\caption{RVs for all single-lined sources. The first and last five rows of the table are presented and the full table is obtainable in the online material.}
\begin{tabular}{r|r|r|r} \\
Target & HJD - 2450000 & RV & $\Delta$RV\\
& \text{[d]} & [km\,s$^{-1}$] & [km\,s$^{-1}$] \\
\hline
WDJ001321.07+282019.83 & 8359.58773 & 30.14 & $\pm$8.56 \\
WDJ001321.07+282019.83 & 8359.59483 & 28.52 & $\pm$6.55 \\
WDJ001321.07+282019.83 & 8359.60215 & 32.48 & $\pm$9.24 \\ 
WDJ001321.07+282019.83 & 8359.60924 & 32.81 & $\pm$7.38 \\
WDJ001321.07+282019.83 & 8359.61633 & 33.83 & $\pm$6.43 \\
\dots & \dots & \dots & \dots \\
\dots & \dots & \dots & \dots \\
WDJ232557.82+255222.39 & 8365.61147 & 10.53 & $\pm$7.46 \\
WDJ233041.67+110206.43 & 8364.60806 & 47.07 & $\pm$3.74 \\
WDJ235313.18+205117.58 & 8360.66009 & 28.75 & $\pm$3.28 \\
WDJ235313.18+205117.58 & 8362.55461 & 26.30 & $\pm$4.79 \\
WDJ235313.18+205117.58 & 8362.57556 & 30.99 & $\pm$2.56 \\
\end{tabular}
\label{tab:DBLsinglelined}
\end{table}


\bsp	
\label{lastpage}
\end{document}